# WILLIAM DAWES: PRACTICAL ASTRONOMY ON THE 'FIRST FLEET' FROM ENGLAND TO AUSTRALIA

**Richard de Grijs**
*Department of Physics and Astronomy, Macquarie University,*
*Balaclava Road, Sydney, NSW 2109, Australia*
Email: richard.de-grijs@mq.edu.au

**and**

**Andrew P. Jacob**
*Sydney Observatory, Museum of Applied Arts and Sciences,*
*1003 Upper Fort Street, Millers Point, Sydney, NSW 2000, Australia*
Email: Andrew.Jacob@maas.museum

**Abstract:** On 13 May 1787, a convict fleet of 11 ships left Portsmouth, England, on a 24,000 km, 8-month-long voyage to New South Wales. The voyage would take the 'First Fleet' under Captain Arthur Phillip via Tenerife (Canary Islands), the port of Rio de Janeiro (Brazil), Table Bay at the southern extremity of the African continent and the southernmost cape of present-day Tasmania to their destination of Botany Bay. Given the navigation tools available at the time and the small size of the convoy's ships, their safe arrival within a few days of each other was a phenomenal achievement. This was particularly so, because they had not lost a single ship and only a relatively small number of crew and convicts. Phillip and his crew had only been able to ensure their success because of the presence of crew members who were highly proficient in practical astronomy, most notably Lieutenant William Dawes. We explore in detail his educational background and the events leading up to Dawes' appointment by the Board of Longitude as the convoy's dedicated astronomer-cum-Marine. In addition to Dawes, John Hunter, second captain of the convoy's flagship H.M.S. *Sirius*, Lieutenant William Bradley and Lieutenant Philip Gidley King were also experts in navigation and longitude determination, using both chronometers and 'lunar distance' measurements. The historical record of the First Fleet's voyage is remarkably accurate, even by today's standards.

## 1 A VOYAGE TO THE ANTIPODES

In a move reeking of desperation, at daybreak on 13 May 1787 a convoy of 11 ships left Portsmouth, England, on an 8-month, 24,000 km undertaking billed as a "voyage to Botany Bay" in New South Wales. Now known as the 'First Fleet', it was a desperate attempt to rid the overcrowded English jails and prison hulks of the 'criminal classes'. As such, Captain Arthur Phillip took charge from his flagship, H.M.S. *Sirius*, accompanied by the H.M.A.T. *Supply* armed tender, the convict transports *Alexander*, *Charlotte*, *Friendship*, *Lady Penrhyn*, *Prince of Wales* and *Scarborough*, and the food and supply transports[1] *Borrowdale*, *Fishb[o]urn* and *Golden Grove* (for an in-depth account, see Pembroke, 2013). Figures 1, 2 and 3 paint a vivid picture of the conditions in which the convicts lived. Figure 1 is a representative example of the numerous prison hulks the English Government operated in the late eighteenth century. Figure 2 shows (left) a chain gang being moved to a new location in ca. 1782 and (right) a longboat carrying convicts to their transport, bound for Botany Bay (early 1800s). Finally, Figure 3 shows the interior of the last Australian prison ship, the prison hulk *Success* (ca. 1926).

The total number of people carried by the First Fleet—convicts, Marines, seamen, civil officers and free people, male and female—will likely remain unknown forever. In fact, it appears from contemporary sources that the convoy's accounting for all of its crew was not done very carefully:

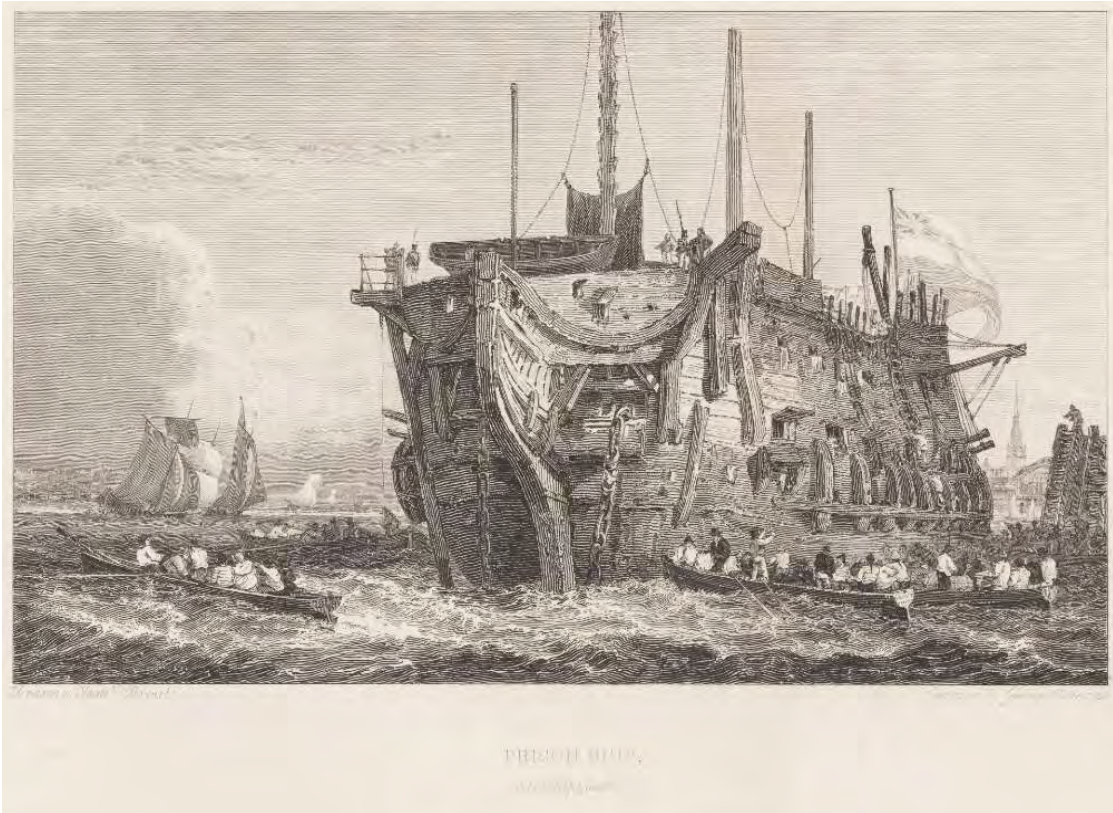

**Figure 1**: Representative example of an English prison hulk at Deptford (1826). Artist: Samuel Prout; engraved by George Cooke (London, Longman & Co.; National Library of Australia; PIC Drawer 3841 #U3905 NK6402).

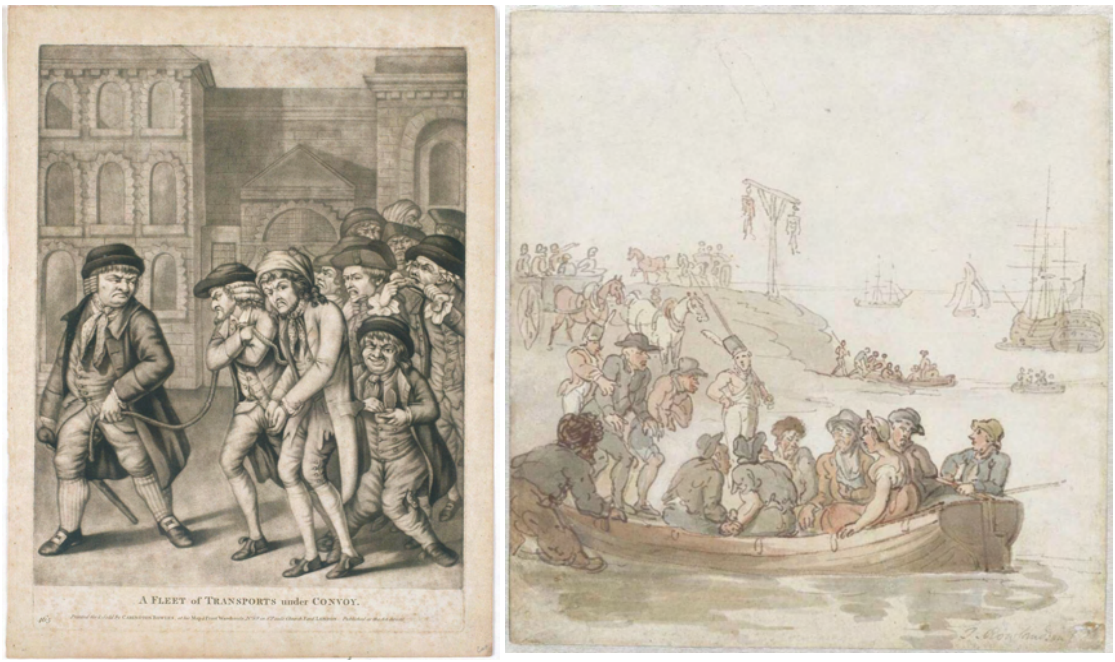

**Figure 2**: (*left*) "A fleet of convicts under convoy", ca. 1782; "Printed for & Sold by Carington Bowles, at his Map & Print Warehouse, No. 69 in St. Pauls Church Yard, London. Published as the Act directs" (Dixon Library, State Library of New South Wales; a128083/DL Pd 789). (*right*) Convicts embarking for Botany Bay, early 1800s. Artist: Thomas Rowlandson (National Library of Australia; PIC Drawer 3842 #T2670 NK228).

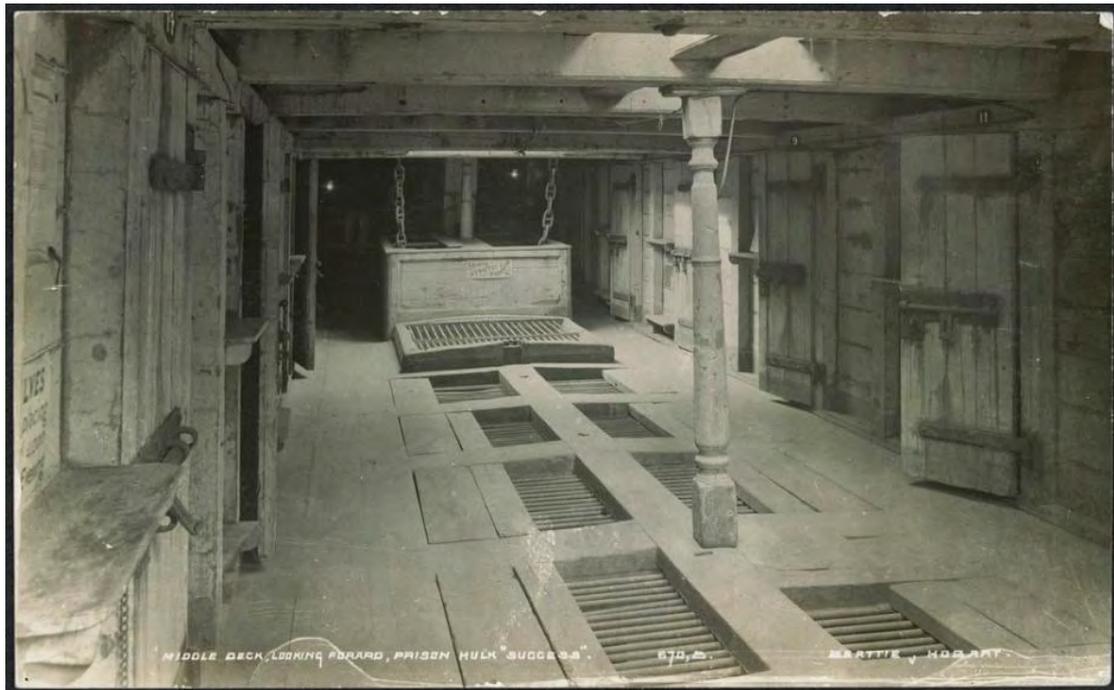
**Figure 3**: Interior of the prison hulk *Success* (middle deck, looking forward), 1926 (State Library of Victoria; Ref. 3007866).

> On the 15th [May 1787] the signal was made for the transports to pass in succession within hail under the stern of the *Sirius*, when, on inquiry, it appeared, that the provost-marshal [the head of the military police] of the settlement (who was to have taken his passage on board the *Prince of Wales*) was left behind, together with the third mate of the *Charlotte* transport, and five men from the *Fishbourn* store-ship: the loss of these [latter] five persons was supplied by as many seamen from on board the [escort ship] *Hyæna*. (Collins, 1798: Section I)

Only for the *Sirius* and *Supply*, neither of which transported convicts, do we know the names of all crew members. Gillen (1989) identified 1420 people who embarked on the First Fleet, with 1373 arriving in the colony after a voyage of 250–252 days, depending on their vessel. The voyage took 48 lives (45 of convicts or their children), while 28 children were born in transit[2] (White, 1790). John White, the First Fleet's principal surgeon (see Figure 4[3]), painted a favourable picture of conditions on board:

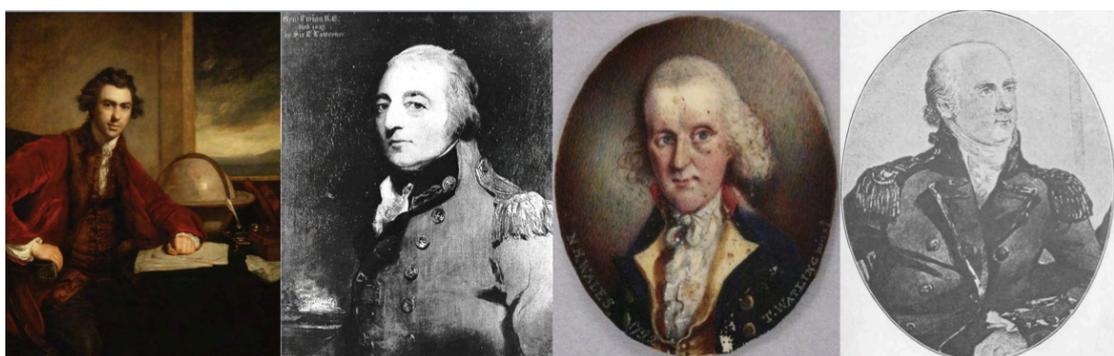
**Figure 4**: Portrait gallery of the main characters described in this article (except for those individuals already included in de Grijs and Jacob, 2021), ordered from left to right by date of birth. Individuals depicted include *(i)* Joseph Banks (1743–1820), *(ii)* William Twiss (1745–1827), *(iii)* John White (1756–1832) and *(iv)* Philip Gidley King (1758–1808). Figure credits: Wikimedia Commons (public domain).

> The newspapers were daily filled with alarming accounts of the fatality that prevailed among us; and the rumour became general, notwithstanding every step was taken to remove these fears, by assurances (which were strictly true) that the whole fleet was in as good a state of health, and as few in it would be found to be ill, at that cold season of the year, as even in the most healthy situation on shore. The clearest testimony that there was more malignity in the report than in the disease, may be deduced from the very inconsiderable number that have died since we left England; which I may safely venture to say is much less than ever was known in so long a voyage (the numbers being proportionate), even though not labouring under the disadvantages we were subject to, and the crowded state we were in.

However, the historical record remains inconclusive as to the total number of people on board. Other accounts report as few as 1044 arrivals, while Governor Phillip's first census of 1788, reported to the Home Secretary, Thomas Townshend, First Viscount Sydney, returned a total white population of 1030 (Macquarie, 1988).[4] Even the number of convicts on board the ships is subject to confusion and debate. Gillen (1989) identified 775 convicts (582 men and 193 women) as having embarked, with 732 eventually landing at Sydney Cove. However, Judge Advocate and chronicler of the colony David Collins (1798) counted 756 convicts across the ships—564 males and 192 females—as well as 13 of their children.

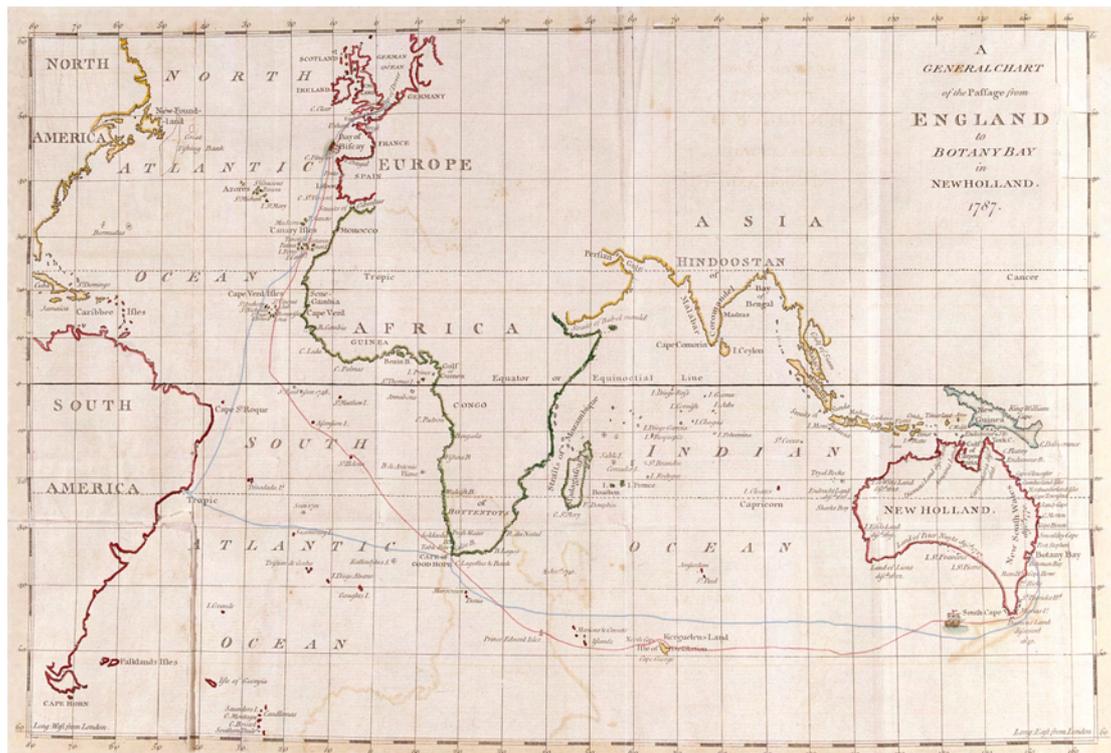

**Figure 5**: "A general chart of the passage from England to Botany Bay in New Holland". Artist: John Andrews (London, John Stockdale). The blue line represents the route taken by the First Fleet (Mitchell Library, State Library of New South Wales; Ref. M2 118/1787/1).

In addition to the convicts and crew, the First Fleet was home to a complement of 160 Marines (see also Laurie, 1988). The 11 ships also carried surgeon-general John White and his servant, on *Charlotte*, as well as six assistant surgeons—two on the *Lady Penrhyn* and one each on *Friendship*, *Scarborough*, *Sirius* and *Supply*. Two surgeon's mates were stationed on the *Sirius*, while the surveyor-general, Augustus Alt, was on board the *Prince of Wales*. A chaplain, the Rev. Richard Johnson, his wife, and a servant and his wife sailed on the *Golden Grove*, whereas the commissary of stores, Andrew Miller, and his servant were on the *Sirius*. The *Sirius* also hosted three additional servants (to Captains Phillip and Hunter). Moreover, a peace officer had embarked on the *Lady Penrhyn* and a civilian settler on the *Alexander*. Finally, 28 soldiers' wives and their 14 children were also on board the convoy[5] (Collins, 1798).

The First Fleet's voyage carried it from England past Madeira, Tenerife (Canary Islands) and the Cape Verde Islands in the North Atlantic Ocean. Then, following the prevailing trade winds and ocean currents, they sailed to Rio de Janeiro and Table Bay at the Cape of Good Hope. Eventually, the voyage took the fleet through the southern Indian Ocean to Van Diemen's Land (present-day Tasmania) and the New South Wales coast (see Figure 5). Navigating long stretches of open ocean, low-lying islands and reefs in coastal areas, extended periods of calm in the tropics (the 'doldrums') where the ships would be subject to a variety of ocean currents, and gale-force winds across the Indian Ocean's 'roaring forties' meant that accurate position determination was of the utmost importance at any time throughout the voyage.

Although chronometers of sufficient operational accuracy to determine one's longitude at sea had become available by the time of James Cook's second and third voyages in 1772–1775 and 1776–1779, respectively, one had to remember to wind them up periodically and to re-calibrate them occasionally given that their rates would decay over time. Recalibration relied upon accurate 'lunar distance' measurements. This required observations of the apparent distances on the sky between well-known bright stars or the Sun and the lunar limb. In turn, these had to be reconciled with tabulated values for those same distances as observed from Greenwich (for a recent review, see de Grijs, 2020). For these reasons, it was common for voyages of discovery to include competent astronomers among their crew (e.g., Macleod and Rehbock, 1988). The First Fleet was no exception (e.g., Saunders, 1990), and so William[6] Dawes (1762–1836) was appointed as the convoy's official astronomer (for a recent review, see also de Grijs and Jacob, 2021).

In the remainder of this article, we will explore in some detail Dawes' background as well as his achievements on board the *Sirius* and, subsequently, the *Supply*. An accessible, relatively brief and popular account of Dawes' contributions and achievements is available at http://www.visitsydneyaustralia.com.au/william-dawes.html. A high-level academic account of the scientific and political context in which the First Fleet's voyage was prepared and subsequently undertaken can be found in Saunders (1990: 24–74). Here, however, we focus in significantly greater detail on Dawes' contributions, while attempting to be as precise as possible in our descriptions and offering quantitative assessments wherever feasible. This article is therefore complementary to Saunders' (1990) more qualitative overview.

Section 2 considers the young Dawes' origins, education and early career as a Marine officer. In Section 3, we will discuss his appointment as the First Fleet's astronomer and the scientific preparations undertaken for their departure to the antipodes (see also Saunders, 1990). Section 4 deals with the astronomical aspects of the First Fleet's voyage, where we will highlight that Dawes was not the only competent astronomer on board. Position determinations based on both chronometer readings and lunar distances were also obtained by Captain John Hunter (Hunter, 1793), second captain of the *Sirius*, and his Lieutenants William Bradley (Bradley, 1786–1792; Collins, 1798) and Philip Gidley King (King, 1787–1790). Finally, in Section 5 we consider the First Fleet's detailed itinerary as recorded in contemporary journals. We provide the most comprehensive compilation of First Fleet geographic position determinations to date by means of an online database. We conclude that the remarkable achievement led by Captain Phillip would not have been possible without a detailed grounding in practical astronomy of key members of his entourage.

## 2 YOUNG WILLIAM DAWES

Perhaps surprisingly given the leading intellectual role he was to be cast into eventually, little first-hand information pertaining to William Dawes survives:

> There is no man among the founders who ought to have given us so much information about himself and his views as Lieutenant Dawes, and there is no man among them who has given so little. He was the scholar of the expedition, man of letters and man of science, explorer, mapmaker, student of language, of anthropology, of astronomy, of botany, of surveying, and of engineering, teacher and philanthropist. The duty to posterity of such a man, in such singular circumstances, was that he should be always writing, and in fact he wrote nothing at all that can now be read. What we know about him we learn wholly from the writings of others; and the scantiness of our information is made the more exasperating by the fact that everything they write about him—apart from two differences of opinion with Governor Phillip, in respect to which a sensible historian will decline to arbitrate—is in tones of high praise both of character and ability. (Wood, 1924: 1)

Currer-Jones (1930), Dawes' great-granddaughter, noted that the Dawes "Family papers, many consisting of Dawes' letters, etc., … had been destroyed after the death of one of his grandsons" and that

> I have heard from Antigua [where Dawes resided from 1813 until his death in 1836] that many of Dawes' papers were destroyed by the terrible hurricane[7] of [21 August] 1871. The utmost was done to decipher the remains of these, but it was found impossible.

The only surviving correspondence by his own hand is composed of his letters to Dr. Nevil Maskelyne, Britain's fifth Astronomer Royal. These are preserved among the papers of the Board of Longitude, which are held at the Cambridge University Library (UK) and freely available online[8] at high resolution.

William Dawes was the first-born son of Benjamin Dawes and Elizabeth Sinnatt of Portsmouth, England. William had four siblings, including his sisters Elizabeth (1764), Mary (1768) and Ann (1769), and his younger brother John (1766). The earliest available record of young William Dawes in the Church of England parish register of St. Thomas à Becket is that of his baptism on 17 March 1762 (Hampshire parish registers, 1653–1875). This suggests that he was in all likelihood born in early 1762, although his exact date of birth has not been recorded.[9] By the late eighteenth century, the time between birth and baptism had slipped from "the Sunday, or other Holy day next after the child be borne, unless upon a great and reasonable cause declared to the Curate" (Anglican Church Prayer Books, 1549 and 1552; cited by Basten, 2015), as commonly practised in the sixteenth and seventeenth centuries, to longer intervals. For instance, one study indicates that in the period 1771–1789, 75% of newborns were baptised within 38 days, while that time interval had increased to 64 days by 1791–1812 (Basten, 2015; see also Berry and Schofield, 1971).

Elizabeth Dawes née Sinnatt had been a widow until she remarried, to Benjamin Dawes, on 15 October 1761 in the same parish church (Blagg and Andrews, 1913: 128). Given that the couple married just five months prior to William's birth suggests a degree of social pressure. His father was employed as clerk-of-works (construction supervisor in charge of quality control) at the Ordnance Office of Portsmouth Naval dockyard (Mander-Jones, 1966).

William would go far in life, despite his fairly humble beginnings. During the early years of the new settlement in New South Wales (1788–1791), Dawes and his newly established observatory (Wood, 1924; de Grijs and Jacob, 2021) were undeniably the intellectual focal point of the colony's small population. His intellectual pursuits and interests clearly showed that he had enjoyed an excellent and thorough education from an early age. If anything, as an adult Dawes developed into a true polymath, proficient in a number of European languages, including French and Latin, as well as in botany, mineralogy, mathematics and maritime navigation. He was also said to be "a tolerably good Astronomer & draws very well" (Bayly, 1786).

His grasp of Latin grammar is demonstrated in his notebooks on the Eora Aboriginal language he would later compile at Sydney Cove (Clarke, 2015). His Latin proficiency is evidenced by his almost casual use of verb paradigms and expressions such as "imperative mood" and "the ablative case" (Steele, 2005). Moreover, Dawes communicated in Latin with the Portuguese astronomers based at the observatory in Rio de Janeiro during the First Fleet's month-long respite there (Dawes, 1787*l*). In addition, shortly after the First Fleet's arrival in New South Wales, Dawes made the acquaintance of the French astronomer Joseph Lepaute Dagelet. The latter was a scientific member of the Lapérouse expedition, which was anchored in Botany Bay between 26 January and 10 March 1788. On 3 March, Dagelet sent Dawes a letter offering him advice on how to set up his observatory (Dagelet, 1788). The letter was written entirely in French. At a later date, Dawes showed further proficiency in French in his notebooks, for instance by referencing "To make or do (faire in French)" (Dawes, 1791: *Notebook* 3, 29).

Although it is unknown where Dawes received his early professional education, historians tend to agree that Dawes most likely attended the Royal Naval Academy in Portsmouth (e.g., Steele, 2005; Crittenden, 2010; Clarke, 2015). His skills in maritime navigation exhibited at a later stage of his career attest to a solid naval education. Nevertheless, Clarke (2015) points out that he continued to perfect those skills as late as 1786 when, prior to the First Fleet's departure, he spent a few weeks under the tutelage of Maskelyne at Greenwich Observatory (Maskelyne, 1786b). Crittenden (2010) verified that Dawes does not appear to have attended either the Naval College at Greenwich or that at Christ's Hospital, the other educational institutions where naval officers were taught the ropes.

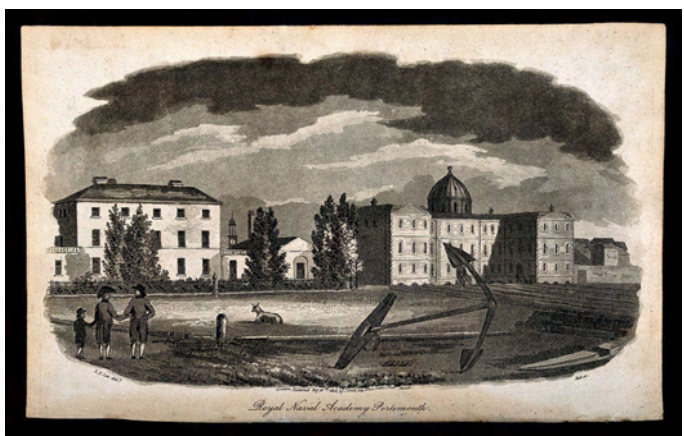

**Figure 6**: Royal Naval Academy, Portsmouth (1806). Artist: Hall, after I. T. Lee (Wellcome Library no. 22276i; Creative Commons Attribution 4.0 International license).

The Portsmouth Naval Academy (see Figure 6) had been established to educate "forty young [13 to 16-year-old] gentlemen, sons of noblemen and gentlemen" (by Order in Council, February 1729) for service in the British Navy. After all, as proclaimed by King William IV, "there was no place superior to the quarterdeck of a British man of war for the education of a gentleman" (Dickinson, 2007: 32). However, that scenario had not developed as initially expected. The offspring of the city's better-off citizens disliked the institution since it reminded them too much of school, with all its rules, responsibilities and restrictions. Upon graduation, the 'Academites' or 'College Volunteers' were equally disliked by the Navy captains required to accept them for sea service, since they lacked the rigorous attitudes of genuine naval officers trained on board as Captain's servants (e.g., Lloyd, 1966; Crittenden, 2010). Moreover, in 1801 Admiral of the Fleet John Jervis, First Earl of St. Vincent, exclaimed that the Portsmouth Naval Academy was "a sink of vice and abomination, which ought to be abolished" (Lloyd, 1952: 472). Given the Academy's bad reputation—"Are you so partial to that seminary as to hazard a son there?" (Jervis, 1801)—by 1773 only half of the approximately 40 places were filled by sons of the city's gentry and aristocrats. Therefore, the Admiralty extended entry to 15 sons of non-commissioned officers, aged 11–17, and waived their fees (Lloyd, 1966: 145).

Under those conditions, and probably aided by his father's position at the Naval dockyard, young William may have been admitted to the Royal Naval Academy. Subjects taught at the time included French, drawing, fencing, use of the firelock and Latin. If indeed the young Dawes was educated there, this would also account for his competence in mathematics, navigation and astronomy. Immersion in the latter subject would serve him well as the official astronomer on the First Fleet. It is indeed highly likely that he received instruction in practical astronomy from William Bayly, who had made a name for himself as astronomer on Captain Cook's second voyage of discovery. Bayly became the Royal Naval Academy's headmaster by February 1785, two years before the First Fleet's departure. As such, he may have been involved in Dawes' advanced training in preparation for his duties during the voyage and, subsequently, in the new colony.

However, despite Dawes' likely education at the Royal Naval Academy, a career in the Navy from an entry-level position as midshipman[10] or Captain's servant was not open to him, given his common, non-aristocratic background. Instead, he enrolled as junior officer in the Marines. Crittenden (2010) suggests that, at one point, he may have considered a career in the Church, which would have been in line with his strong religious leanings and high morals (e.g., Crittenden, 2010; James, 2012; Clarke, 2015). For instance, following the First Fleet's arrival in New South Wales, Lieutenant Daniel Southwell, Master's mate on the *Sirius*, wrote to his mother,

> [Dawes] is a most amiable man, and though young, truly religious, without any appearance of formal sanctity. He is kind to everyone; but I am speaking of his many affabilities to myself, which are such that more could not be looked for from a relation. He has a great share of general knowledge, studious, yet ever cheerful, and the goodness of his disposition renders him esteemed and respected by all who know him. (Southwell, 1790)

However, as in the Navy, clergy in the Church were also vying for positions and promotions. Thus, qualifications came second to the old boys' network (e.g., Crittenden, 2010), where the local nobility dominated. Crittenden (2010) muses that Dawes may have heard John Wesley (the founder of the Anglican Church's Methodist movement) preach at one of his prayer meetings in the open air, attempting to bring religion back to the masses. This would have appealed to young William. Instead, however, he opted for the advantage of the home court (after all, he grew up on the Navy dockyard) and joined the Marines.

Upon graduation from the Royal Naval Academy in 1779, aged 17, Dawes was gazetted (commissioned) as a second Lieutenant in the British Marine Corps.[11] He joined the 32nd Company of the Portsmouth Division (Wright, 1927; Steele, 2005). Great Britain was at war. The American War of Independence (1775–1783), also known as the American Revolutionary War, had been triggered by the British Government's plans to tax the colonies. This led to a general revolt, united in their opposition—"no taxation without representation" (in Parliament). What had started as an uprising, with expedient attempts of suppression by the British Army and Navy, turned into a full-scale war by 1779. The French sailed in support of the American colonies, with Spain and the Dutch Republic following suit.

William Dawes was posted to H.M.S. *Resolution* on 5 April 1780 (Ship's muster, H.M.S. *Resolution*, The National Archives, Admiralty 36/8709), propelling him into the rough and tumble world of the warring nations. The *Resolution* was a third-rate[12] 'ship of the line' equipped with 74 guns. After Dawes had been posted, the company sailed promptly to the Americas, and so Dawes saw action perhaps sooner than he had expected. The *Resolution* was part of the British fleet commanded by Rear Admiral Sir Thomas Graves. They engaged the French fleet under Rear Admiral François Joseph Paul, Comte de Grasse, in battle off Chesapeake Bay (Maryland) in early September 1781. Dawes was wounded during the main battle of 2 September, although probably not very seriously given that he continued to serve on the *Resolution* (Nicolas, 1845: 116–117).

Next, despite the *Resolution*'s new deployment as part of Admiral George Rodney's British fleet, she saw renewed hostile engagement with de Grasse's fleet in the Battle of the Saintes (9–12 April 1782), off the coast of Dominica (West Indies). Whereas the Battle of Chesapeake had been a strategic victory for the French, the Battle of the Saintes (also known as the Battle of Dominica) was a clear British victory. It offered a significant morale boost for the British troops at home and overseas. Upon arrival back at their home port on 23 October 1782, Dawes left the *Resolution* (Ship's muster, H.M.S. *Resolution*, The National Archives, Admiralty 36/8713). He next enlisted in the 11th Company at Sheerness. His subsequent service returned him to American waters, while he was also engaged in patrolling the southern North Sea on the sloop H.M.S. *Merlin* (Ship's muster, H.M.S. *Merlin*, The National Archives, Admiralty 36/10463; Gillen, 1989: 101) between 1783 and 1784. On the *Merlin*, he was put in charge of a small Marine contingent, which continued until he left service on the *Merlin* on 15 May 1784. He was subsequently assigned to the Marine barracks in Portsmouth. There, he pursued improvements of his surveying and engineering skills while also establishing himself as a competent astronomer.

**3 OFFICIAL 'FIRST FLEET' ASTRONOMER**

Most accounts of Dawes' involvement in the First Fleet's voyage simply refer to him having volunteered for service as a Marine officer, without further consideration. This cavalier approach to referencing probably dates back as far as Watkin Tench's 1827 appeal to Earl Bathurst, Secretary of State for the Colonies. Tench requested that Dawes be awarded additional financial compensation[13] for his work in the new settlement: "… Your Lordship's Memorialist [Dawes] was among the Officers who volunteered their services to go to Botany Bay, in the Year 1786, and the only one from the Portsmouth Division" (Wright, 1926).

However, the sequence of events that eventually led to Dawes' appointment as second Lieutenant of the Marines on the *Sirius* was complex, with multiple obstacles that had to be overcome. Dawes' good fortune seems to have hinged in large part on an early intervention by Bayly, already in his role as Master of the Portsmouth Royal Naval Academy. On 8 August 1786, Bayly wrote to Joseph Banks, then-President of the Royal Society of London, to recommend Dawes' "great desire to go" with the First Fleet, extolling the Marine's numerous useful technical and linguistic skills (Clarke, 2015). He assured Banks that all of his lavish praise of Dawes was "strictly true".

**Figure 7**: Captain Arthur Phillip's signed receipt for the Board of Longitude's instrumentation and books on loan (Board of Longitude papers, Cambridge University).

Bayly soon developed into Dawes' main early patron. Since he had served as assistant to Maskelyne (Howse, n.d.), Bayly had a direct line to the nation's top scientist. He exploited his direct access to lobby the Astronomer Royal on Dawes' behalf.[14] In turn, Maskelyne wrote to Banks on 17 October 1786 that Dawes was "well versed in most kinds of astronomical observations" (Maskelyne, 1786a), thus supporting the Board of Longitude's recommendation of Dawes' candidacy as astronomer on the voyage. He followed up on 8 November (for a transcript, see de Grijs and Jacob, 2021), further emphasising Dawes' suitability for the appointment.

Dawes received additional support from another local Portsmouth contact, Captain William Twiss of the Royal Engineers. The latter was in charge of construction at the Naval dockyard and port. On 24 October 1786, Twiss sent a letter to Brook Watson, then a Member of Parliament and a Director of the Bank of England, recommending Dawes' services for the voyage to Botany Bay. Watson forwarded Twiss's letter to Evan Nepean, Under-Secretary of State for the Home Office with responsibility for naval and political intelligence (Twiss, 1786).

Here the timeline becomes confusing. We learn from a letter of 25 October 1786 from Phillip Stephens, Secretary to the Admiralty, to Lieutenant-General Smith, Commandant of the Marines, Portsmouth Division, that Dawes was granted leave to sort out his "private affairs" (Stephens, 1786). This included further instruction in and development of his technical skills as a practical astronomer at the Royal Greenwich Observatory (Maskelyne, 1786b; Clarke, 2015). This timing of Dawes' leave for purposes of reviewing his technical astronomy skills appears odd, given that Maskelyne's proposal to the Board of Longitude to appoint Dawes as the First Fleet's astronomer was not discussed until 14 November that year. The Board of Longitude's minutes of 14 November 1786 record that Maskelyne, *ex officio* Board member,

> ... represented to the Board that Mr Dawes Lieutenant of Marines on board His Majesty's Ship the *Sirius* which is going with the Convicts to Botany Bay on the Coast of New Holland, is desirous of making useful Nautical & Astronomical observations in his passage thither, … (Board of Longitude, 1786)

This is also the first time that the establishment of an observatory in the new colony is addressed (see de Grijs and Jacob, 2021):

> … and during his stay there, if he could be allowed to use some Instruments belonging to the Commissioners of the Longitude, to enable him to do so; And the Astronomer Royal having at the same time laid before the Board a List of Instruments and Books proper to lend Mr Dawes to enable him to make the said Observations, and at the same time [Maskelyne] informed the Board that Mr Dawes was capable of making proper use of them. The Board took the same into Consideration and came to a Resolution to lend the Instruments and Books undermentioned for the purposes aforesaid, but being of the Opinion that if they were put into the Charge of Captain Philip [*sic*], the Commander of the *Sirius*, greater care would be taken of them, than could be done by Mr Dawes, they directed that the said Instruments and Books should be accordingly delivered into the Charge of Captain Philip, and that he should be requested to give his Attention to the care and preservation of them.

An entry in William Bradley's journal of 5 December 1786 provides details about the suite of instrumentation that would be available to Dawes on the voyage (see also Laurie, 1988; Saunders, 1990: 41–44, 49–51; de Grijs and Jacob 2021):

> The Board of Longitude furnished the following instruments for the use of the voyage and the new colony, An astronomical Quadrant of one foot radius, A 3½ feet treble object glass achromatic telescope by Dollond, with a wire Micrometer for measuring diff.$^{ce}$ of right ascension & declination, also a Micrometer with oblique wires and a Quadrant fixed to it with a moveable short Telescope to take the distance of any object (nearly) from bright fixed stars. A night glass. An Astronomical clock. A Journeyman clock. An Alarum clock. An old sextant by Ramsden. A portable Barometer & two thermometers. These instruments Captain Phillip gave a receipt for [see Figure 7], promising to return them to the Board (the dangers of the sea and other unavoidable accidents excepted) at his return on a receipt from such Officer as may supercede [*sic*] him in the command. Lieut. Dawes of the Marines a volunteer for the Botany Bay Detachment having been introduced to Dr Maskelyne the Astronomer Royale [*sic*], he was acknowledged a proper person to make such observations on shore as might be judged of use. (Bradley, 1786–1792: 3)

Dawes only owned two instruments himself (Clarke, 2015). Therefore, Maskelyne's intervention with Banks on the Marine's behalf (Maskelyne, 1786a) secured the success of the astronomical mission. In addition to the instruments lent by the Board of Longitude,

Maskelyne was also instructed to supply Phillip with Larcum Kendall's 'K1' chronometer, an exact copy of John Harrison's Longitude Prize-winning watch 'H4', which had served Captain Cook so well on his second and third voyages to the Pacific. It is clear, however, that the Board valued the hierarchy on board the fleet. It was thus decided to lend the scientific instruments in the care of Captain Phillip, despite Maskelyne's assurance that Dawes was "capable of making proper use of them". This decision was most likely driven by Phillip's more spacious and secure accommodation on board the *Sirius*. Nevertheless, although Dawes was initially pleased that the instruments were Phillip's responsibility, he expressed some concern about the inconvenience of the distance from his berth to Phillip's cabin:

> In your letter of the 15 Inst.$^t$ you inform[ed] me that Capt.$^n$ Phillip will give you a Receipt for all the Instruments and Books belonging to the Board, which I received from you, and will return me my Receipt, &c; on the first perusal of the foregoing it did not occur to me that any Inconveniency could arise from such a Circumstance, but was rather pleased at the Idea of the whole Charge of valuable Books and Instruments being taken off my Shoulders; but on revolving it in my Mind, I cannot but imagine that unless the *Nautical Almanac*s and *Requisite Tables* be entirely at my own disposal and for my own peculiar Use, numberless Inconveniences and Delays will arise: my Cabin is at a considerable Distance from Capt.$^n$ Phillip's and to be obliged on every Occasion to send to him for a *Nautical Almanac*, at which Time perhaps he may be using it and not able to spare it, must be exceedingly troublesome and disagreeable to both of us. (Dawes, 1786)

Since access to the *Requisite Tables* was deemed essential, in the same letter Dawes requested an additional copy from Maskelyne, who agreed to supply one. In addition to Dawes' concern about access to the astronomical instruments during the voyage, Phillip did not readily inform the astronomer when developments occurred. For instance, five cases containing the scientific instruments Dawes had requested were delivered to the *Sirius* on 5 December 1786 (Bradley, 1786–1792: 3), but in his letter of 8 February 1787, Dawes (1787d) informed Maskelyne that he still had not been advised as to the crates' contents. His patience was put to the test until 20 March of that year before he was able to inspect Phillip's receipt, dated 16 February 1787 (Dawes, 1787e).

When he eventually managed to inspect Phillip's receipt, Dawes (1787e) noticed that Maskelyne's "small telescope with oblique wires in the focus" was not among the listed instruments. He was, however, rather keen to acquire that telescope, since

> … it [is] very probable that I may be frequently detached some little Distance from the fix'd Observatory, when with such a Telescope as that, and a proper Stand to it I should be able to make the necessary Observations in any place with the help of a good Watch, of which I believe there will be several in [*sic*] the Ship, one of which I could borrow, should the Board of Longitude's prove insufficient; however if the night Telescope be furnished with oblique Wires; that will answer the Purpose effectually I should suppose. (Dawes, 1787e)

Meanwhile, Dawes' personal circumstances, where he had to satisfy multiple paymasters simultaneously, had the potential to derail the entire astronomical project. Indeed, despite the strong support from Maskelyne, Banks and the Board of Longitude, Dawes' anticipated roles as Marine, the First Fleet's astronomer-designate and the new colony's scientist in charge of establishing a permanent observatory were at risk. Whereas the home nation's scientific establishment might have found their man, the military hierarchy did not necessarily agree. Dawes' appointment was in some jeopardy. Although it is often suggested that he had volunteered for the mission too late to be appointed as member of the squadron's four shore-based companies of Marines (Eldershaw, 1972: 40; Tench, 1979: 77), Saunders (1990: 48) suggests that he was assigned by the Admiralty to the Marine complement aboard the *Sirius* in good faith, with the expectation that he "might have leisure independent of other avocations" (Lord Howe, 1787) to pursue his scientific tasks. However, his shipboard allocation came with inherent disadvantages compared with membership of the general Marine contingent—in terms of his wages, opportunities and even career advancement.

Whereas members of the main, shore-based Marine contingent were offered a year's advance subsistence pay, the shipboard complement did not qualify for these benefits. Clearly unhappy, Dawes (1787a) appealed to Admiral Richard Howe, First Earl Howe, First Lord of the Admiralty and Chair of the Commissioners of the Longitude, arguing that he had been unfairly excluded. Lord Howe did not indulge the young Lieutenant; he responded tersely that the Marine complement had signed on for a fixed term in New South Wales, while the *Sirius* contingent might be redeployed at any time:

> The Officers of the Detachment are to remain on Garrison Service for the Term already communicated to them. – The party embarked in the *Sirius*, on the contrary, is liable to be ordered home with the Ship at the expiration of 12, 18 or any greater Number of Months as the Exigencies of the Service may require: and both is, & could only be considered in all cases, as similar in Circumstances with every other marine Detachment ordered for foreign Stations. (Howe, 1787)

This advice concerned Dawes greatly: as a ship's officer he was subject to the whims of his commanding officer and as such he could be sent anywhere the ship would need to go, at any time. In the late eighteenth century, British Marines were soldiers enlisted as part of the Naval establishment, and so they were expected to live and remain on board the ship that served as their duty station. Moreover, while attached to the ship's contingent, Dawes was not automatically granted leave to go ashore to construct the observatory he had been tasked to establish. Both conditions could potentially jeopardise the fledgling astronomical endeavour in the new settlement. However, Lord Howe continued that he did not consider Dawes' orders to construct an observatory of paramount importance:

> The Astronomical Observations there might be opportunity to take are of very inferior moment and may be totally omitted, without prejudice to the intention of the Armament. Wherefore, if you are desirous of being relieved in the *Sirius*, and the time will admit of it with reasonable attention to the convenience of the Officer next in tour of Duty at the Division, I don't at present see any Objection to the concurrence of the Board in such request as you may see fit to make on that occasion.

Indeed, Lord Howe, not known as a pushover under any circumstances, called the young Marine's bluff.

Nevertheless, Dawes persisted in his efforts to enact a change in his circumstances, hoping for Maskelyne to exert his influence: "I think it will be allowed that my Situation is truely [*sic*] hard and distressing" (Dawes, 1787b). Dawes argued that he was in danger of losing his seniority in his shipboard role, to which he forcefully objected in a letter to Maskelyne of 25 January 1787:

> I wrote to Capt.$^n$ Phillip Yesterday informing him of what had passed relative to my appointment to the *Sirius* &c. & requesting he would endeavour to get me removed from her to the shore on the first Vacancy which may happen in the Detachment; this would not prevent my going out in the Ship. — I am very little solicitous about getting or saving Money, my greatest Fear is that my being on board the *Sirius* will prevent my promotion Case [illegible] Vacancies should happen above me, as I cannot possibly succeed to a superior Rank without being absolutely in the Detachment where the Vacancies (if any) will happen; I would not lose one Day's Rank for Tons of Gold & Diamonds & this may probably be the Case unless I get appointed to the Shore Detachment. I should think the loss of Rank the greatest Misfortune that could possibly befal[l] me; in short I would much sooner lose a Limb & would almost as soon lose my Life. (Dawes, 1787c)

Therefore, Dawes promptly requested, although unsuccessfully, that he be reassigned to the shore-based contingent at Botany Bay. Dawes' insistence that he did not care much about being compensated financially does not fully ring true, however. This issue touches on the ultimate motivation for the men to volunteer for the Botany Bay expedition. Whereas courage, a sense of adventure, an inquisitive mind and an interest in exploration were undoubtedly important attributes of the volunteers' characters (e.g., Wood, 1924; Clarke, 2015), their realistic alternative looked rather bleak, with limited opportunities for career progression and retention on half pay only (e.g., Wright, 1926; Clarke, 2015). It is, therefore, not surprising that Dawes did not raise the issue of his shipboard assignment again in his subsequent correspondence with Maskelyne. He appears to have accepted his appointment on the *Sirius* without further complaints,[15] although he requested to be considered for the first available shore-based vacancy (Dawes, 1787b).

Dawes spent his remaining time prior to their departure from England on calibrating his instruments, preparing an observing plan for their sojourn at the Cape of Good Hope and determining the accuracy of the chronometer's clock rate. The latter aspect caused him great concern. In collaboration with Bayly, he carefully compared the chronometer's rate with the clocks at the Portsmouth Royal Naval Academy and found worrying irregularities. These appeared to depend on the extent to which the timekeeper's spring was wound. In addition, the sextant he had been issued with developed a fault (see also Saunders, 1990: 50), possibly because of an accidental drop, and so he requested a replacement instrument through Maskelyne's assistant,[16] George Gilpin. This expenditure was authorised by the Board of Longitude at its meeting of 3 February 1787. The new sextant, supplied by the leading instrument maker Jesse Ramsden, cost £13 7*s*. 0*d*.; the cost of repairs to the other instruments came to £44 13*s*. 0*d*. (Board of Longitude, 1787).

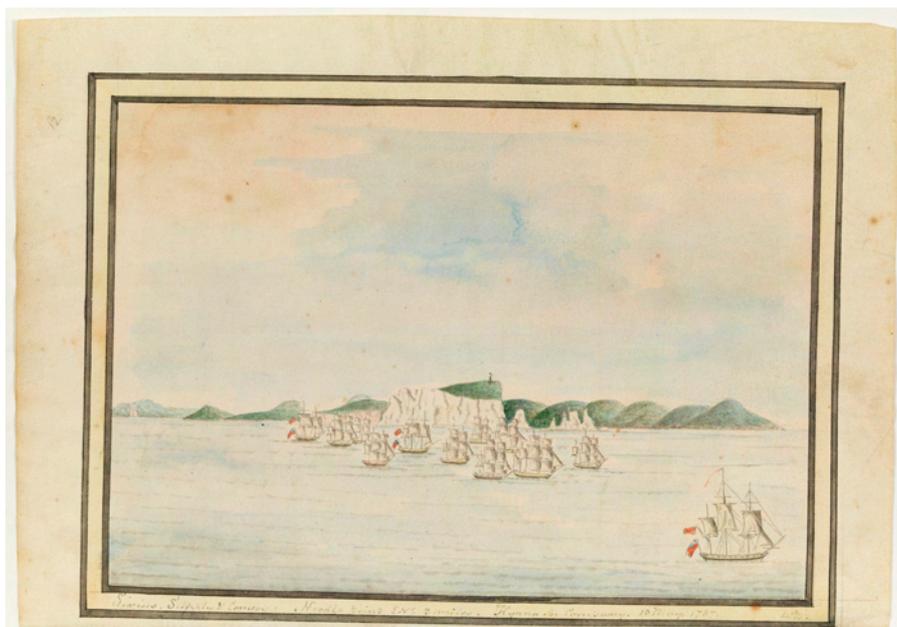

**Figure 8**: Departure of the First Fleet, passing the Isle of Wight's 'Needles', 13 May 1787. Artist: William Bradley, *A Voyage to New South Wales* (ca. 1802) (Mitchell Library, State Library of New South Wales; Ref. 412997).

Shortly before their departure, Dawes told Maskelyne that Captain Phillip had arrived on board the *Sirius* and, more importantly, that he was favourably disposed of Dawes' astronomical endeavours (Dawes, 1787f). Bradley (1786–1792: 11) highlighted Phillip's apparent interest in the scientific aspects of the voyage at that time and their final preparations:

> May 7.[th] Captain Phillip arrived at Portsmouth. He brought with him a timekeeper made by Mr. Kendal[l] and a sextant, both furnished by the Board of Longitude for the use of the voyage. The timekeeper was sent by an officer to the Royal Academy at Portsmouth, and left in charge of Mr. Bayley [*sic*] the head master. … The 11.[th] Waiting only for a wind to carry the ships to sea, an officer was sent to bring on board the timekeeper, the rate of its going determined by Mr. Bayley. The 3 days which he had it, it lost 1"38 per day.

It later transpired that this period of appeasement between Dawes and Phillip, which appears to have lasted until Dawes allowed K1 to run down following their departure from the Cape (see Section 4),[17] was the only time that the men were on cordial terms.[18] Dawes had apparently been warned about Phillip's allegedly difficult personality:

> I am well convinced of the Propriety of paying every proper respect and Attention to Capt.[n] Phillip, and shall on all Occasions endeavour to make my self as serviceable to him as possible, and am well aware that in Case of difference of Opinions on any subject: if cool reasoning and fair Argument have not the desired Effect, a more harsh Method of proceeding whether with Superiors or Inferiors cannot by any Means do Good but may do much Harm. (Dawes, 1787g)

**4 *EN ROUTE* TO BOTANY BAY**

Upon the First Fleet's departure from English shores (see Figure 8), Dawes' most important daily task consisted of maintaining the flagship's main chronometer, K1, given its importance for the convoy's navigation to Botany Bay. Phillip had issued a set of meticulous orders detailing the protocol that was to be adhered to. Bradley's notes (Bradley, 1786–1792: 11) specify that the 'timekeeper' had to be rewound every day at noon in the presence of Phillip with either Hunter or Dawes (it required two keys), as well as the guard on duty outside Phillip's cabin. Subsequently, the day's readings had to be confirmed by the officer of the watch:

> The precautions necessary to prevent the timekeeper from being let down were ordered by Captain Phillip who, with Captain Hunter or Mr. Dawes, were always to be present at the winding it at noon. And it was ordered to be the duty of the lieutenant who brought 12 o'clock to see it done and the officer who relieved him was not to take charge of the deck until he was informed that it was done. The sentinel at the cabin door was also ordered to plant himself inside the cabin, on hearing the bell ring at noon, and not to go out to be relieved until he was told, or saw, that the timekeeper was wound up by one of the officers. (Bradley, 1786–1792: 11)

However, K1 was not the only portable timepiece on board the *Sirius*. Dawes had been lent an Ellicott watch[19] by Maskelyne (Board of Longitude, 1786; Laurie, 1988), whereas Hunter owned a "reliable chronometer" made by John Brockbank; we will discuss Hunter's use of his own chronometer in navigating the southern Indian Ocean in more detail below.

Dawes' other scientific duties involved recording daily observations of the weather, including the temperature and air pressure (e.g., McAfee, 1981; Ashcroft, 2016). He also routinely recorded the Earth's gravity on the basis of pendulum timings (e.g., Forbes, 1975: 172), compass deviations as the voyage progressed (e.g., Hunter, 1793: 23) and their position at sea using either the timepiece or lunar distance measurements, or ideally both, to derive their actual longitude. Collins (1798: Section II) explains,

> The longitude, when calculated by either altitudes of the sun, for the time-piece, (of Kendal[l]'s constructing, which was sent out by the Board of Longitude,) or by the means of several sets of lunar observations, which were taken by Captain Hunter, Lieutenant Bradley, and Lieutenant Dawes, was constantly shewn to the convoy, for which purpose the signal was made for the whole to pass under the stern of the *Sirius*, when a board was set up in some conspicuous part of the ship with the longitude marked on it to that day at noon.

Hunter further clarified that their longitude determination "was [routinely] marked with chalk in large characters on a black painted board, and shewn over the stern to the convoy" (Hunter, 1793: 36).

As a matter of protocol, Dawes compared his results with those of Hunter and Bradley, and he sent his observations back to Maskelyne at every possible opportunity (e.g., Dawes, 1787h). The first such opportunity occurred on 20 May 1787 by means of the homeward-bound frigate H.M.S. *Hyæna* under Captain Michael de Courcy (Coursey). The *Hyæna* had escorted the convoy from Spithead roads (anchorage) off Portsmouth to a point some 300 miles west of the island of Ushant (Ouessant, France), at "Latitude 47° 57½ N, Longitude 12° 14½ W by Timekeeper this Day at Noon" (Dawes, 1787g).

Hunter, clearly impressed by Dawes' careful work, referred to Dawes as "a young gentleman very well qualified for such a business, and who promises fair, if he pursues his studies, to make a respectable figure in the science of astronomy" (Bach, 1968: 13). Hunter was a proficient astronomer and navigator in his own right, as we will see below. As a case in point, consider the following passage from his journal shortly after their departure:

> On the 15th [May 1787], at sun-set, the Start Point bore north-east half east by compass, distant seven or eight leagues: at noon on this day (which finishes the nautical and begins the astronomical day) the longitude, by account [based on dead reckoning], was 5° 01' west of the meridian of Greenwich, and by a time-piece made by Mr. Kendal[l], with which the Board of Longitude had supplied us, it was 4° 59' west; ... our lunar observations, which we never failed to make at every opportunity, constantly confirmed the truth of the watch. (Hunter, 1793: 5, 15)

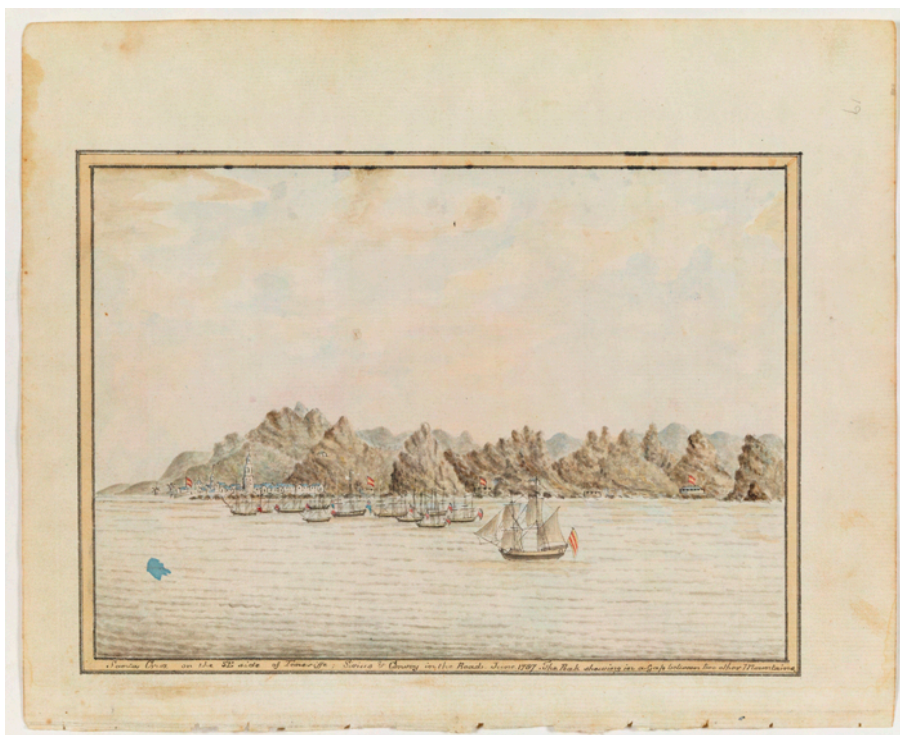

**Figure 9**: "Santa Cruz on the SE side of Teneriffe; *Sirius* & Convoy in the Roads. June 1787. The Peak Shewing in a Gap between two other Mountains". Artist: William Bradley, *A Voyage to New South Wales* (ca. 1802) (Mitchell Library, State Library of New South Wales; Ref. 412997).

Having passed close by Madeira and the Savage Islands (Salvage Islands, Ilhas Selvagens), the convoy reached Santa Cruz bay at Tenerife in the Canary Islands on 3 June 1787 (see Figure 9), "after a Passage, for the most Part pleasant and agreeable" (Dawes, 1787h). Dawes continued, "we made as much use of the Time Keeper & our Sextants as the Weather would permit us, we only miss'd getting the Longitude by the T.K.$^r$ [Time Keeper] two Days during the whole Passage". Although Dawes was keen to make observations from Santa Cruz during their week-long stay there, Phillip refused permission to unload the pendulum clock and the quadrant, since they were stored deep in the hold, below the convoy's stores of bread:[20]

> Immediately on our Arrival at the Place I applied to Capt.$^n$ Phillip to get the Clock and Quadrant on Shore; but he informed me that it was absolutely impossible to make any satisfactory Use of them at this Place, as he is determined to get away on Thursday [7 June 1787] if possible ... besides which, the Motion of the Ship has caused the Bread to cover the Cases so entirely that they could not be got at without much difficulty & the Risk of getting a good deal of the Bread damaged; but on our Arrival at Rio [de] Janeiro, we shall not only have more Time, but the Bread will then be so much reduced as to obviate any difficulty in that Respect. ... I shall attend most assiduously to the T.K.$^r$ [Time Keeper] while we remain at this Place & hope with the Sextant to be able to attain to almost as much accuracy as if I had the Astronomical Quadrant on Shore. (Dawes, 1787h)

The ships' sojourn at Tenerife was predominantly meant to restock their wine, water and other provisions for the next leg of the voyage down to the South Atlantic. Dawes wrote two letters to Maskelyne from the Santa Cruz roads, both dated 5 June 1787, which included a range of longitude and latitude measurements (Dawes, 1787h,i).

It would take another eight weeks following the convoy's departure from Tenerife on 10 June before they arrived at St. Sebastian (Rio de Janeiro) by 5 August 1787 (see Figure 10). They had been forced to abandon their plans for a brief stop in the Cape Verde archipelago because of adverse wind conditions at the St. Jago (Santiago) roads. Since a number of officers on the First Fleet kept detailed accounts of the voyage in their personal journals, we will reproduce some of the most pertinent passages related to the convoy's navigation and the astronomical endeavours here:

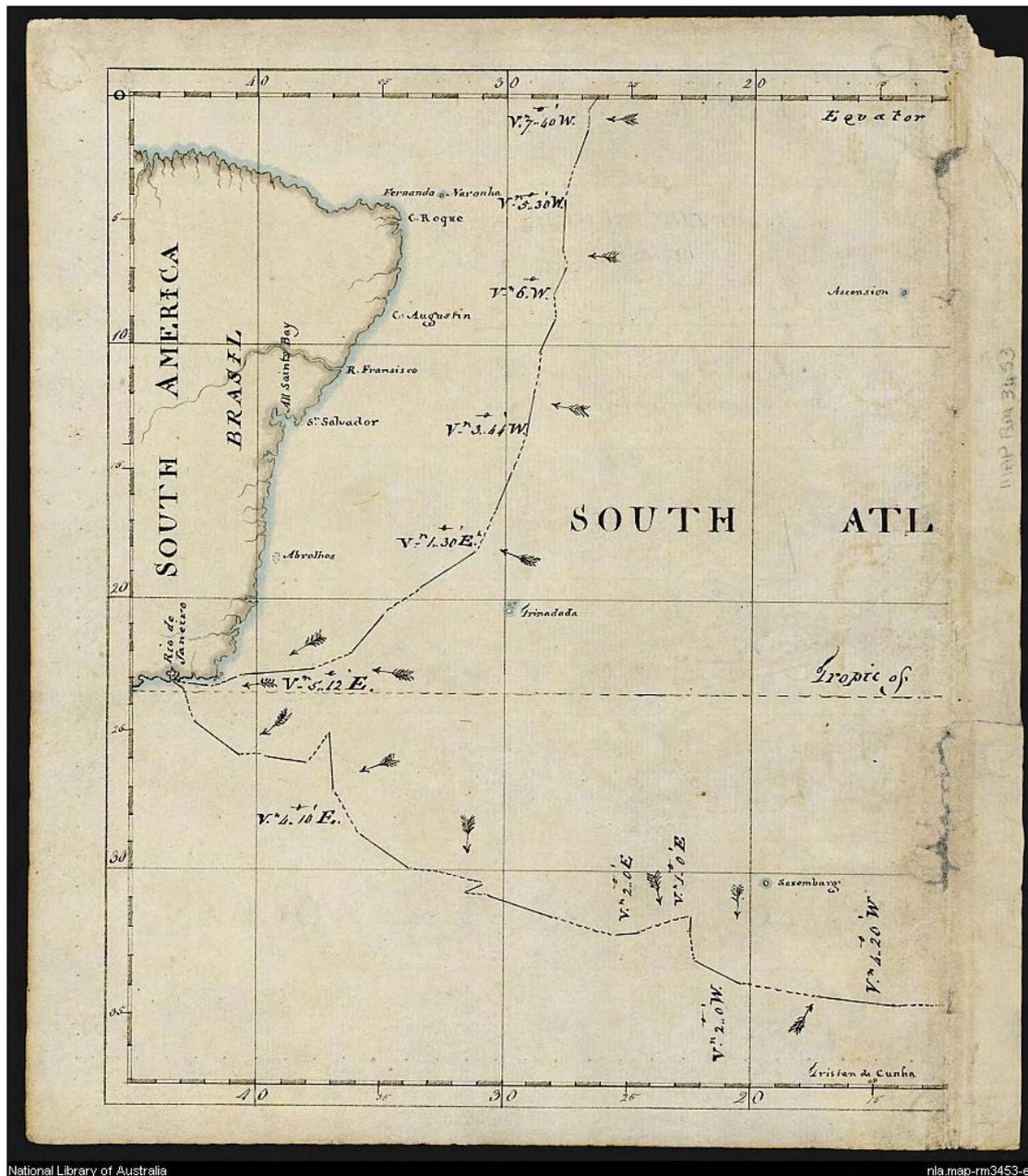

**Figure 10**: Chart of the South Atlantic showing the route of the *Sirius*, 1787–1788 (George Raper; National Library of Australia; Ref. nla.map-rm3453).

> On Thursday the 2d of August we had the coast of South America in sight; and the head-land, named Cape [Cabo] Frio, was distinctly seen before the evening closed in. Our timepiece had given us notice when to look out for it, and the land was made precisely to the hour in which it had taught us to expect it.[21] It was not, however, until the evening of the 4th that we anchored within the islands at the entrance of the harbour of Rio de Janeiro.
> …
> On the morning after our arrival the intendant of the port, with the usual officers, repaired on board the *Sirius*, requiring the customary certificates to be given, as to what nation she belonged to, whither bound, the name of her commander, and his reason for coming into that port; to all which satisfactory answers were given; and at eleven o'clock the day following Captain Phillip, accompanied by the officers of the settlement, civil and military, waited upon Don Louis Vasconcellos, the viceroy of the Brazils, at his excellency's palace, who received them with much politeness, readily assenting to a tent being pitched on shore for the purpose of an observatory; ... (Collins, 1798: Section II)

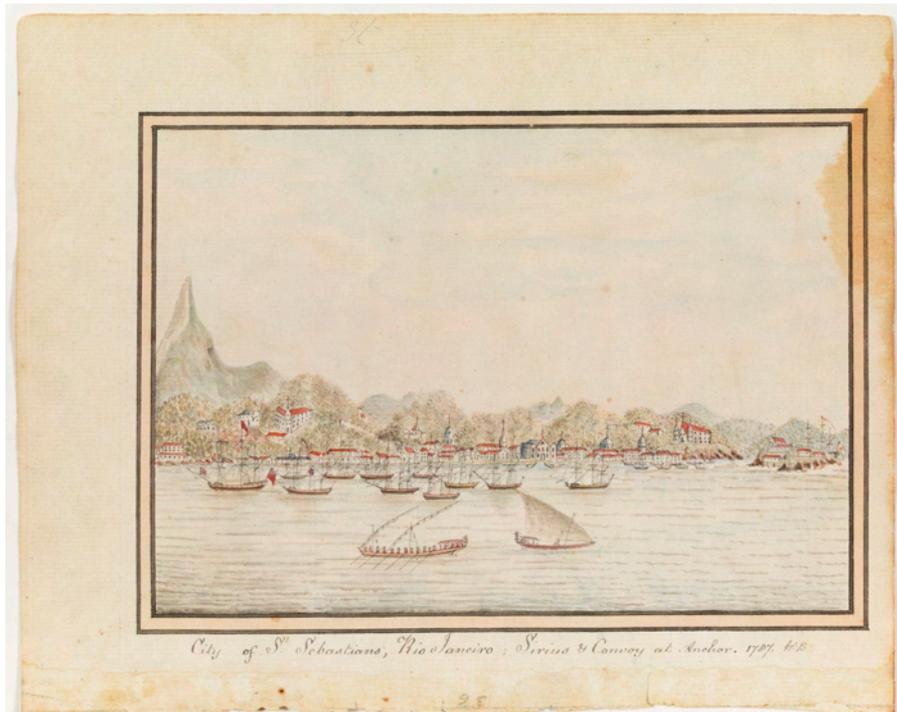

**Figure 11**: "City of St. Sebastian, Rio Janeiro: *Sirius* & Convoy at Anchor. 1787" Artist: William Bradley, *A Voyage to New South Wales* (ca. 1802) (Mitchell Library, State Library of New South Wales; Ref. 412997).

The day after their arrival at St. Sebastian (see Figure 11), Dawes was indeed busy. He wrote an urgent letter to Maskelyne requesting that some of his spoilt books be replaced at the first opportunity. The letter was taken back to Europe by a passing Portuguese ship bound for Lisbon:

> My Cabbin [*sic*] being under Water (the greater part of it) I have frequent Seas coming into the Scuttle in consequence of which & the constant Use I have made of them, the *Requisite Tables* are much the Worse ... (Dawes, 1787i; see also Dawes, 1787k for a repeat request)

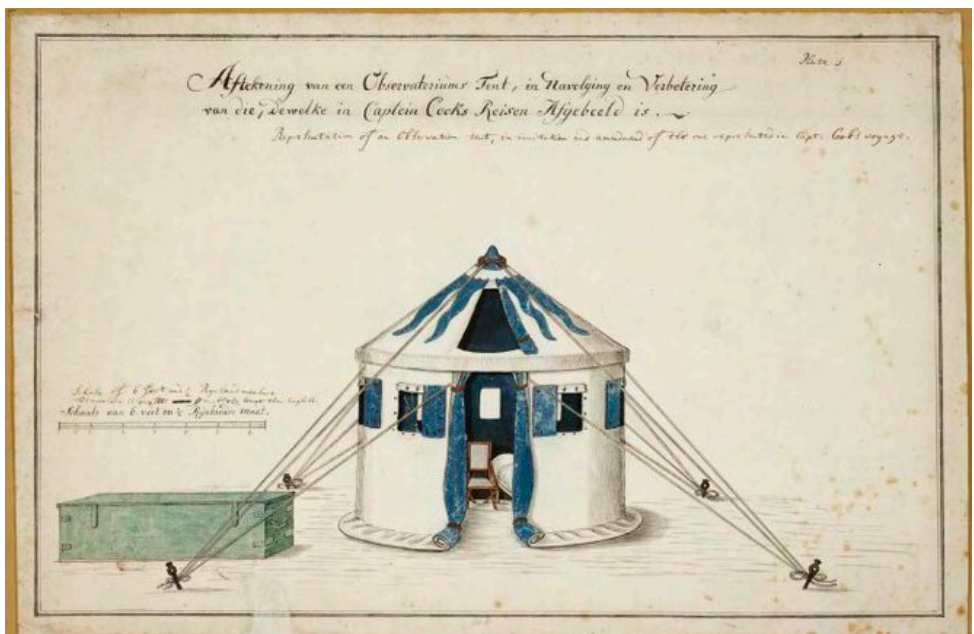

**Figure 12**: Typical example of the type of tent observatory Dawes would have pitched on Enchados island in the harbour of Rio de Janeiro (Artist unknown: *Aftekening van een observatoriums tent in navolging en verbetering van die, dewelke in Captain Cook's reisen afgebeeld is*. Plate 1, ca. 1780? Ref.: B-091-007. Alexander Turnbull Library, Wellington, New Zealand. Ref. /records/22320145).

On account of his earlier service in the Portuguese navy (e.g., Pembroke, 2013), Captain Phillip was received and treated with the highest regard, and so his officers were given unprecedented freedom to move around the city. This preferential treatment also included permission to establish a temporary tent observatory (for a typical example, see Figure 12). Dawes (1787l: 269r) requested it to be pitched on the harbour island of Enchados (Ilha das Enxadas), close to the city:

> 9.th [August 1787] Captain Phillip and Mr. Dawes went to look at a room which had been granted for the purpose of making the necessary observations for the timekeeper etc., but it not being thought proper for the purpose, Captain Phillip obtained leave from the Viceroy to make use of an Island (Enxadas) lying to the NE of the town, which spot being approved, a tent was erected on the 11th and on the 12th, the instruments landed and fixed under the direction of Mr. Dawes. (Bradley, 1786–1892: 36)

> During our stay here, we were permitted to erect a tent on the island of Enchados, (a small island about a mile and a half farther up the harbour than where we lay with the ships,) for the purpose of landing a few of the astronomical instruments which were necessary for ascertaining the rate of the time-keeper; they were put under the charge and management of Lieutenant William Dawes of the marines ... (Hunter, 1793: 20–21)

> The weather was rather unfavourable, during the time the instruments were on shore for ascertaining the rate of the time-keeper, but as constant attention was paid, every opportunity that offered was made use of, and the watch [rate] was found to be 2"-27 which is near a second more than was its rate at Portsmouth. (Hunter, 1793: 21)

Despite the inclement weather (see also King, 1787–1790), for the next month Dawes, assisted by two young sailors from the *Sirius* (Collins, 1798: Section II), proceeded apace with his observational programme. His observations included measurements of the ephemerides of Jupiter's satellites, and particularly of the eclipses of the planet's third largest moon, Ganymede (Dawes, 1787l: 271v). He made sure to express his appreciation of Phillip's encouragement and support, particularly of the fact that Phillip himself had covered the costs associated with establishing the temporary observatory. The Portuguese authorities provided guards to ensure that no one would disturb the observatory's business (Dawes, 1787l)—or perhaps to keep an eye on their progress: "The 13.th [August 1787] The Timekeeper was sent to the tent and all boats belonging to the transports strictly forbidden landing on that island" (Bradley, 1786–1792: 36).

Dawes had meanwhile also ascertained the rates of his clocks and felt confident of their performance. In his letter of 3 September 1787 (Dawes, 1787i), he sketched the set-up of his astronomical and journeyman clocks (see Figure 13), placed back-to-back, explaining the need for his rather intricate support system:

> The Earth being very hard it was not in my Power to get the Clock Frame set up 'till Wednesday Morning, when the Earth was rammed exceedingly well all round it, so that the Frame was as firm & steady as possible at and near the Surface of the Earth, but on trying it at the Top, I found it required but little Force to shake it considerably, so that I thought necessary to get four large Pickets & as many Shoars made … the Pickets being driven into the Earth very firmly at about three Feet distance from the Angles of the Frame & the Shoars placed as represented in the Plan sketched below …

Indeed, Dawes appears to have been a careful and thorough observer (Bradley, 1786–1792; White, 1790; Tench, 1979: Introduction):

> If Circumstances permit I always take 6 at least Observations at one Time, whether Altitudes for the Longitude by Time Keeper or Distances for the Lon[gitude] by the Lunar Method. I divide them into two Setts [*sic*] of three each and work the Mean of each three separately & then take a mean of the Result of each Sett; by this Means I think Mistakes may almost always be detected. The Means of each Sett have in general agreed exceedingly well. The greatest Difference has only amounted to about 4" of Time, but they have almost always agreed within 2" of each other. – I shall not remit in the least from endeavouring by a great Number of Observations to get the true Rate of the Time Keeper as well as the Longitude of this or any other Place as nearly as possible by Distances, Occultations &c.[a] (Dawes, 1787i)

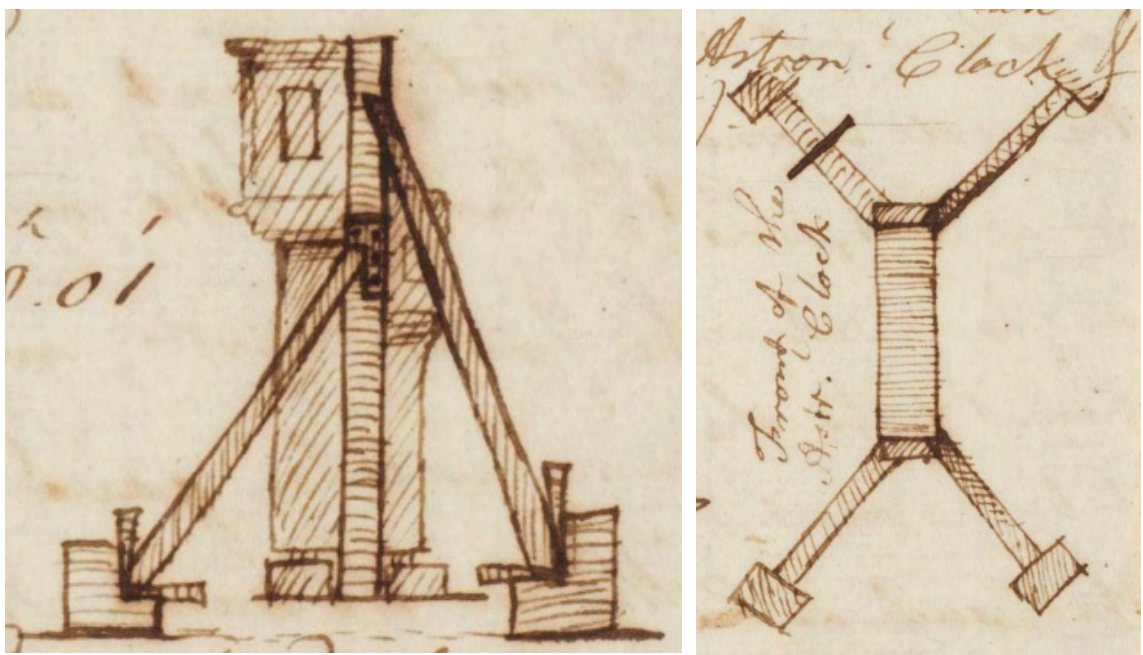

**Figure 13**: Dawes' sketches of the setup of his astronomical and journeyman clocks, placed back-to-back and supported by large struts, in his temporary observatory at Enchados island in Rio de Janeiro (Dawes, 1787g).

| City of St. Sebastian. | Deg. | Min. | Sec. |
|---|---|---|---|
| Latitude | 22 | 54 | 13 south. |
| Longitude, deduced from our time-keeper of the meridian of Greenwich, and which agrees with that laid down in the new requisite tables, but which certainly are not correct. | 42 | 44 | 00 west |
| Longitude, determined by two astronomers sent from Portugal for that and other purposes | 43 | 18 | 45 west. |
| Longitude, by an eclipse of Jupiter's third satellite, taken by Lieutenant Dawes, on the island Enchados | 43 | 19 | 00 west. |
| Longitude, by a mean of several distances of ☉ and ☽ taken by me at the outer anchorage | 43 | 11 | 15 west. |
| Ditto, by Lieutenant Bradley | 43 | 33 | 00 west. |

**Figure 14**: Comparison of the First Fleet's longitude determinations of the city of St. Sebastian (Rio de Janeiro) as recorded by Hunter (1793: 50).

As a case in point, he determined the longitude of Cape Frio to within a third of a degree of its actual longitude (equivalent to approximately 19 nautical miles at the latitude of Rio de Janeiro; Laurie, 1988). Rio de Janeiro served as a very useful calibration benchmark for the convoy's position determinations: see Figure 14. To his credit, Dawes diligently compared the performance of his own instruments with those at the local observatory, and he would do the same upon their arrival at the Cape of Good Hope. On 3 September, he wrote to Maskelyne,

> ... on Saturday the 11.$^{th}$ [August 1787] ... I went on board to inform C.P. [Captain Phillip] that the instrument might be sent on shore that afternoon, which he immediately gave directions for, and I accordingly took them on shore with me the same afternoon. On Monday Morning I set up the Quadrant, and in the evening, the clock being set up ...

> ... On the 17.$^{th}$ [August 1787] in the forenoon I was invited by two astronomers employed here by the Court of Portugal to settle the limits between the Portuguese & Spanish Settlements in this Country. They were brought to the Tent by Capt.$^{n}$ Phillip & I immediately foresaw that I might possibly get some observations from them which would be acceptable to you, & therefore determined if possible to return their Visit before our quitting this Place: they admired the Quadrant & took notice how equal the Beats of the Clock were. – On the 1.$^{st}$ of Sept.$^{r}$ [1787] I went on shore to return their Visit and found one of them at home who very willingly allowed me to copy the Observations contained in the two Sheets accompanying this ... The Gentleman told me he had made many meteorological observations ...

The Portuguese astronomers Dawes referred to in his letter to Maskelyne were based at the city's observatory. Watkin Tench provides a little more detail about the key issue of contention in Brazil at the time:

> Among other public buildings, I had almost forgot[ten] to mention an observatory, which stands near the middle of the town, and is tolerably well furnished with astronomical instruments. During our stay here, some Spanish and Portuguese mathematicians were endeavouring to determine the boundaries of the territories belonging to their respective Crowns. Unhappily, however, for the cause of science, these gentlemen have not hitherto been able to coincide in their accounts, so that very little information on this head, to be depended upon, could be gained. How far political motives may have caused this disagreement, I do not presume to decide; though it deserves notice, that the Portuguese accuse the Abbee de la Caille, who observed here by order of the King of France, of having laid down the longitude of this place forty-five miles too much to the eastward. (Tench, 1979: Chapter V)

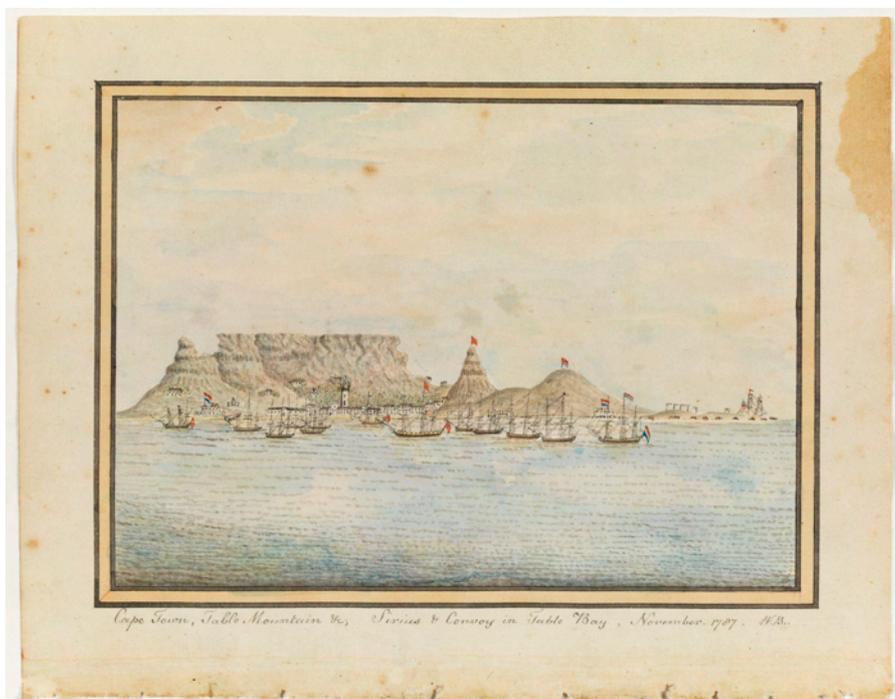

**Figure 15**: "Cape Town, Table Mountain &c; *Sirius* & Convoy in Table Bay, November. 1787" Artist: William Bradley, *A Voyage to New South Wales* (ca. 1802) (Mitchell Library, State Library of New South Wales; Ref. 412997).

From both Rio de Janeiro and the Cape of Good Hope, Dawes encouraged Maskelyne to keep in touch with the local astronomers:

> ... he [Senhor Bento Sanchez d'Orta] would with pleasure communicate to you if you should think proper to enter into a correspondence with him ... I also informed him of the comet expected in 1789 [see below] which he promised to pay as much attention to as he could. (Dawes, 1787l: 273v)

Almost a month after their arrival in Rio de Janeiro, the convoy left, on 4 September 1787, on their way to the Cape of Good Hope. They arrived in Table Bay on 13 October that year (see Figure 15):

> This run, from about lat[itude] 22° south, long[itude] 43° west of London, to lat[itude] 34° south, long[itude] 18° east of London, a distance of about four thousand miles, was performed in thirty-nine days: for having left Rio on the 4$^{th}$ of September, 1787, on the 13$^{th}$ of October the ships came to anchor in Table Bay. (Phillip, 1790a: 43)

Dawes informed Maskelyne that "the whole Fleet is in perfect good Order & every body exceedingly healthy; only abt [*sic*] 15 Deaths have happened since our departure ..." (Dawes, 1787m). Although Captain Phillip had initially planned to stay at the Cape for just a week, the convoy did not depart until a month had passed, on 12 November 1787. Yet, and to Dawes' great disappointment, Phillip once again did not permit him to disembark his instruments, this time citing safety concerns and fearing that the instruments might sustain damage in the rush of loading cattle and provisions (e.g., Dawes, 1787m; Laurie, 1988; Clarke, 2015). And so Dawes' carefully prepared observing programme for the Cape came to nothing.

Shortly after the convoy's arrival in Table Bay, Hunter questioned the accuracy of the timepiece:

> By altitude taken this morning [15 October 1787] for the time-keeper, it appear'd that we had not had sufficient time at Rio de Janeiro for ascertaining the true rate of the watch's going, having determined what we have allowed this passage, viz. 2"-33 from a very few observations, and those not to be relied on, the weather having been very unfavourable; for, by the difference of time between the meridian of Rio de Janeiro and the Cape, both which places are well determined, the watch has lost at the rate of 3"-17, which we shall hereafter allow to be the true rate; and as a proof of that having been really its rate all along, by allowing it from the time of our leaving Portsmouth, until our arrival at Rio de Janeiro, we shall have the longitude of that place 43° 33' 30" west of the meridian of Greenwich, which is 45' 45" to the westward of that laid down in the new *Requisite Tables*, and which agrees very nearly with the observations made on the spot. (Hunter, 1793: 30)

Meanwhile, as the ships were being stocked one final time with provisions, stores and livestock for the new colony, Dawes was introduced to Colonel Robert Jacob Gordon, commander of the Dutch troops at the Cape, and a great host to the visiting convoy:

> … a gentleman whose thirst for natural knowledge amply qualified him to be of service to us, not only in procuring a great variety of the best seeds and plants, but in pointing out the culture, the soil, and the proper time of introducing them into the ground. (Collins, 1798: Section II)

Gordon owned a Ramsden quadrant, an Arnold watch "in gold, price 60£ spared to him by Capt.$^n$ Cummings in the East India Service" (Dawes, 1787m), various telescopes, almanacs and astronomical tables. However, he remained a novice in terms of having done any observing other than taking latitudes (Laurie, 1988). Dawes convinced him to help with their efforts to observe the comet that Maskelyne had predicted would reoccur in the southern hemisphere in 1788, possibly as early as 1 January of that year (Maskelyne, 1786c; Saunders, 1990; for a recent discussion, see de Grijs and Jacob, 2021): "We have ever since [having been invited to stay with Gordon] been fully employed in putting things in a train for observing the comet which is expected about this time next year" (Dawes, 1787m). Enthusiastically, Gordon ordered a telescope from Dollond through Maskelyne, which Gordon arranged to be paid for by the Dutch consul in London. He also initiated a subscription to *Taylor's Logarithms*, and asked Maskelyne to provide him with the necessary instructions to embark on an observational programme aimed at finding Maskelyne's comet.

Dawes was clearly enamoured with the man and saw his potential as an ally in the southern hemisphere:

> This gentleman is rema[rk]able for his great love of science and Attention to persons e[m]ployed in promoting it ... he has some knowledge of astronomy and is more determined to employ the greatest part of his leisure time to the study of it ... it occurred to me, that it would be doing you the greatest pleasure possible, to embrace so fair an opportunity of adding one to the small number of observers in the southern hemisphere ... (Dawes, 1787m)

Dawes' next letter to Maskelyne was sent from New South Wales.

At this point, we should probably pause for a moment and consider what the next and final leg of the voyage meant to the settlers. The mixed emotions that may have been felt by a number of the men are perhaps best reflected by David Collins' eloquent reminiscence upon their departure from Table Bay:

> It was natural to indulge at this moment a melancholy reflection which obtruded itself upon the mind. The land behind us was the abode of a civilized people; that before us was the residence of savages. When, if ever, we might again enjoy the commerce of the world, was doubtful and uncertain. The refreshments and the pleasures of which we had so liberally partaken at the Cape, were to be exchanged for coarse fare and hard labour at New South Wales. All communication with families and friends now cut off, we were leaving the world behind us, to enter on a state unknown; and, as if it had been necessary to imprint this idea more strongly on our minds, and to render the sensation still more poignant, at the close of the evening we spoke a ship from London. The metropolis of our native country, its pleasures, its wealth, and its consequence, thus accidentally presented to the mind, failed not to afford a most striking contrast with the object now principally in our view. (Collins, 1798: Section II)

The First Fleet left the Cape on 12 November 1787, a day later than anticipated because of adverse winds. Dawes excitedly told Maskelyne in his letter of 9 November 1787 (Dawes, 1787m) that Gordon had supplied him with "several books & instruments which will be of service ... among which are several spare barometer tubes". Meanwhile, Hunter had determined the longitude of Cape Town on a number of occasions during their visit, which both Hunter and Collins recorded in their respective journals (although the actual values differ slightly):

> During the time we lay in this bay [Table Bay], I took a considerable number of lunar observations, by a mean of which I make Cape Town, in longitude 18° 24' 30" east of the meridian of Greenwich: latitude observed in the bay, 33° 55' south, ... (Hunter, 1793: 32)

> Before we quitted the Cape Captain Hunter determined the longitude of the Cape-town in Table-bay to be, by the mean of several sets of lunar observations taken on board the *Sirius*, 18° 23' 55" east from Greenwich. (Collins, 1798: Section II)

Less than a fortnight after their departure, on 25 November "being then only 80 leagues eastward of the Cape" (Phillip, 1790a: 49), Captain Phillip, Lieutenant Gidley King and Dawes[22] transferred from the *Sirius* to the faster *Supply*, aiming to reach Botany Bay sooner than the main convoy (see also Hunter, 1793: 32):

> For several days after we had sailed, the wind was unfavourable, and blowing fresh, with much sea, some time elapsed before we had reached to the eastward of the Cape of Good Hope. On the 16th, Captain Phillip signified his intention of proceeding forward in the *Supply*, with the view of arriving in New South Wales so long before the principal part of the fleet, as to be able to fix on a clear and proper place for the settlement. Lieutenant Shortland was at the same time informed, that he was to quit the fleet with the *Alexander*, taking on with him the *Scarborough* and *Friendship* transports. These three ships had on board the greater part of the male convicts, whom Captain Phillip had sanguine hopes of employing to much advantage, before the *Sirius*, with that part of the fleet which was to remain under Captain Hunter's direction, should arrive upon the coast. This separation, the first that had occurred, did not take place until the 25th, on which day Captain Phillip went on board the *Supply*, taking with him, from the *Sirius*, Lieutenants King and Dawes, with the time-keeper. (Collins, 1798: Section II)

In Hunter's account of the event, we learn that Phillip also took along with him "several sawyers, carpenters, blacksmiths and other mechanics" (see also Tench, 1979: Chapter VII). In addition, he specifically addresses the fate of Kendall's timepiece:

> On the 25th [November 1787], being in latitude 38° 40' south, and longitude 25° 05' east, Captain Phillip embarked on board the *Supply*, in order to proceed singly in that vessel to the coast of New South Wales, where he made sure of arriving a fortnight or three weeks before us, as some of the convoy sailed very heavy; he took with him from the *Sirius*, Mr. Philip Gidley King, second lieutenant, and Lieutenant Dawes, of the marines, who had hitherto kept an account of the time-keeper, which he also took with him; ... (Hunter, 1793: 32)

Phillip's transfer to the *Supply* and his decision to take the K1 chronometer with him on the faster vessel caused significant consternation among the officers remaining behind (e.g., Saunders, 1990: 62, 64). Bowes-Smyth, the convoy's Surgeon-General, referred to that decision as "a mere abortion of the brain, a whim which struck him at the time as the sequel will evince" (Choat, 2020: 61). Meanwhile, King assessed the *Supply* "much too small for so long a voyage, which, added to her not being able to carry any quantity of provisions, and her sailing very ill, renders her a very improper vessel [*sic*] for this service" (King, 1787–1790: 513), and thus he thought that Phillip "flattered himself" (King, 1787–1790: 531) in believing that he would be able to make up a fortnight with respect to the remainder of the fleet. Captain James Campbell of the *Lady Penrhyn* was incensed and expressed his dismay that Phillip was "indifferent about everything but his own safety" (Campbell, 1788: 98) in deciding to proceed with "the Don Quixote scheme of separating our little Fleet – leaving them to work their way through an immense sea but little known, and to which all were strangers" (Campbell, 1788: 96).

And alas, the transfer of K1 from the *Sirius* to the *Supply* indirectly led to accidental negligence when, on 18 December 1787, the timepiece was not rewound at noon and instead ran down as it was forgotten: "[On 25 January 1788,] We received the Timekeeper from the *Supply* where it had unfortunately been let down on the passage to this place" (Bradley, 1786–1792: 64). It took a significant number of lunar distance measurements and lengthy calculations before Dawes was able to reset and recalibrate the clock mechanism (Bradley, 1786–1792: 6; King, 1787). Dawes later explained to Maskelyne the circumstances leading to the accidental stoppage:

> Whenever there is any thing to be got out of the hold, it is a very awkward rather dangerous thing to go from the quarter deck to the cabbin [*sic*] it was owing to this that on the 18th Dec.[r] [1787] the time-keeper was let down, as Capt.[n] Phillip could not get down at noon to wind it up & it was not thought of afterwards by anyone till near six o'clock at which time it had been down above an hour; however the next day I got some exceeding[ly] good altitudes, from which the longitude was carried on, so that we were only liable to the inaccuracy of the log for two days. (Dawes, 1788)

However, when K1 stopped, rewinding it did not fully return it to its former operation; rewinding it caused the timekeeper's rate to change as well. Dawes' careful observations led him to conclude that the watch had been idle for a total of 1 hour 9 min 44 sec (Morrison and Barko, 2009; see also King, 1787–1790). It took until September 1788, well after their arrival in Sydney Cove, before Dawes was able to fully recalibrate K1 (Morrison and Barko, 2009).

Meanwhile, with K1 on the *Supply* and the rest of the convoy split into two smaller squadrons, the ship's captains were left to their own devices. Captain Hunter of the *Sirius* was the most qualified astronomer remaining with the main fleet. Both Hunter and Collins provide detailed accounts of their progress across the southern Indian Ocean.

> After the time-keeper was taken from the *Sirius*, I kept an account of the ship's way by my own watch, which I had found for a considerable time, to go very well with Kendal[l]'s; I knew it could be depended on sufficiently to carry on from one lunar observation to another, without any material error; for although its rate of going was not so regular as I could have wished, yet its variation would not in a week or ten days have amounted to any thing of consequence; it was made for me by Mr. John Brockbank, of Cornhill, London, upon an improved principle of his own. The lunar observation, which I never failed to take [at] every opportunity, and which Lieutenant Bradley also paid constant attention to, gave me reason to think, by their near agreement with the watch, that it continued to go well. (Hunter, 1793: 33)

For three successive days [from 1 December 1787] both Mr. Bradley and myself had a variety of distances, by which our account seemed to be very correct. I now determined (if I could avoid it) never to get to the northward of latitude 40° 00' south, and to keep between that parallel and 43° or 44° south. After the 3d [December 1787], I found, by altitudes taken for the watch, that we went farther to the eastward than the log gave us, and no opportunity offered for getting a lunar observation to compare with it until the 13th, when both Mr. Bradley and I got several good distances of the ☉ [Sun] and ☽, by which our longitude was 70° 22' east, by the watch 70° 07' east, and by account [dead reckoning] 67° 37' east. On the 14th, the weather being very clear, we had another set of distances, which gave our longitude 73° 06' east, by the watch 73° 09' east, and by account 70° 34' east. Again, on the 15th, I observed with two different instruments, one by Ramsden, and the other by Dollond, and the results agreed within ten miles of longitude; the mean was 75° 18' east, by the watch 75° 16' east, and by account 72° 49' east. Mr. Bradley's mean was also 75° 18' east: so that, as I have already observed, the ship seemed [to be] gaining on the account; but there was no reason to believe, that in the middle of this very extensive ocean we were ever subject to much current: I therefore attribute this set to the eastward, to the large following sea, which constantly attended us, since we had taken a more southerly parallel. (Hunter, 1793: 33–34)

In early January 1788, the astronomers on the *Supply* and the *Sirius* both sighted New Holland. Dawes, on the *Supply*, determined the longitude of Tasmania's South Cape on 3 January. It compared favourably with Cook's surprisingly accurate 1777 measurement of 146° 07' 00" East (i.e., within 10 km of the present-day value; Hunter, 1793: 39; Morrison and Barko, 2009), which had also been obtained by reference to Kendall's K1 chronometer.[23] This allowed Dawes to refine the recalibration of K1 by adjusting the timepiece negatively by 3 min 21.7 sec (King, 1788).

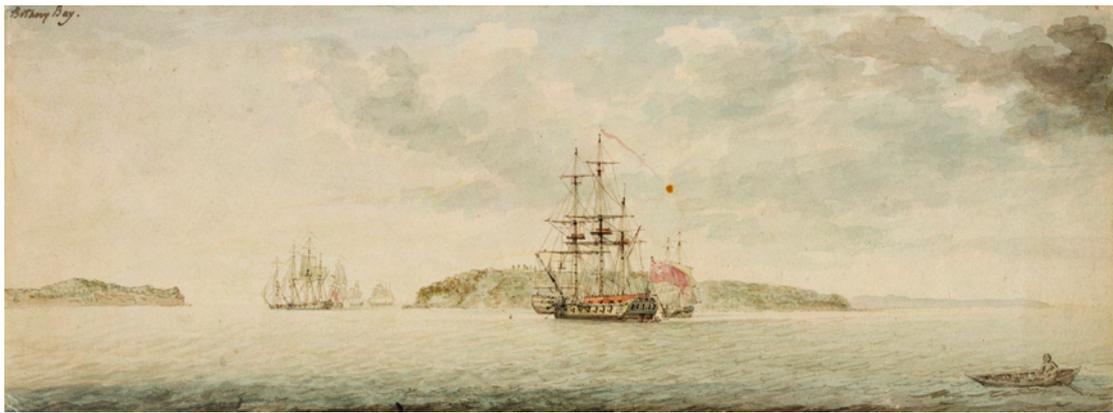

**Figure 16**: Arrival of the *Supply* at Botany Bay, 1788. Artist: Charles Gore (Mitchell Library, State Library of New South Wales; Ref. DG VIA/8).

Following a beautiful display of the aurora australis on 6 January 1788 (see also Bradley, 1786–1792: 51; Tench, 1979: Chapter VII),

This night the aurora austreales were very bright, of a beautiful crimson colour, streaked with orange, yellow, and white, and these colours were constantly changing their places: the highest part was about 45° above the horizon, and it spread from south by east to south-south-west. (Hunter, 1793: 37)

… we learn from Collins (see also Southwell, 1788) that …

… a lunar observation taken at ten o'clock of the forenoon of Monday the 7th, the fleet was then distant seventeen leagues from the South Cape of New Holland; and at five minutes past two in the afternoon the signal was made for seeing the land.

Nothing could more strongly prove the excellence and utility of lunar observations, than the accuracy with which we made the land in this long voyage from the Cape of Good Hope, there not being a league difference between our expectation of seeing it, and the real appearance of it. (Collins, 1798: Section III)

The *Supply* reached Botany Bay on 18 January 1788 (see Figure 16), followed by the *Alexander*, *Friendship* and *Scarborough* on the 19th, while the remainder of the convoy, including the *Sirius*, arrived at their destination on 20 January 1788 (see Figure 17).

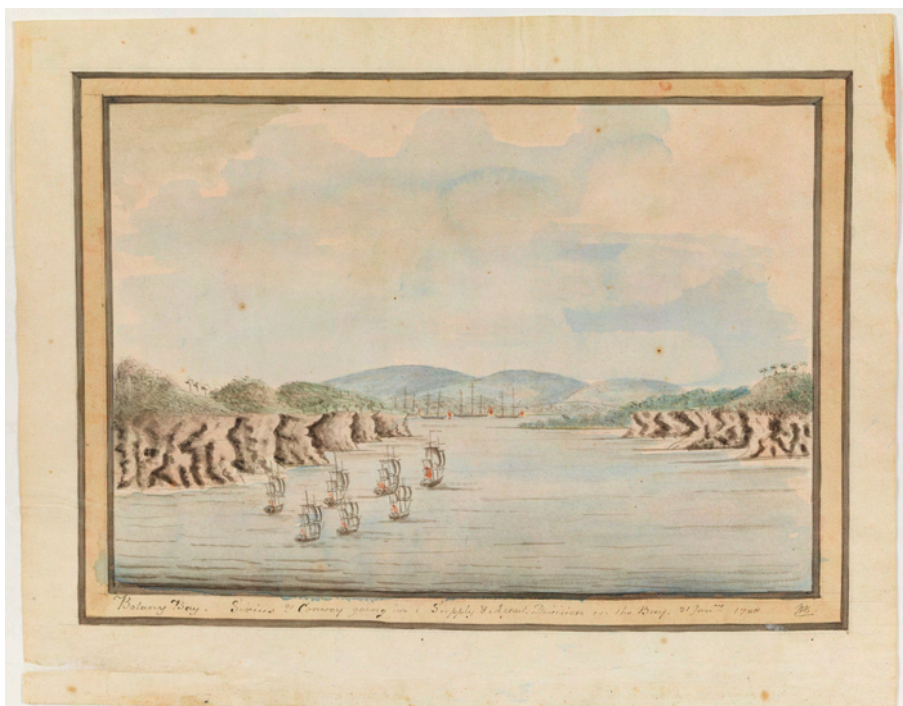

**Figure 17**: "Botany Bay. *Sirius* & Convoy going in: *Supply* & Agents Division in the Bay, 21 Janry 1788". Artist: William Bradley, *A Voyage to New South Wales* (ca. 1802) (Mitchell Library, State Library of New South Wales; Ref. 412997).

**5 CONCLUDING THOUGHTS**

During the 250–252 day voyage of the First Fleet, latitudes and longitudes were determined as often as possible, subject to the prevailing weather. Latitudes were most easily measured, sometimes even in slightly overcast conditions. Determination of the Sun's altitude at its meridian passage would often suffice. Longitude determination based on reading off the time on either K1 or Hunter's personal timepiece was intrinsically more complicated. It required altitude determinations of celestial objects—the Sun, the Moon or bright stars—combined with their predicted locations tabulated in the relevant *Nautical Almanac* and the *Requisite Tables*.

As far as we are aware, a full and comprehensive account of the First Fleet's positions as the voyage progressed is not generally available (but see Gergis et al., 2010). Dawes included a table of the *Sirius*' longitudes and latitudes between 14 May and 2 June 1787 in his letter to Maskelyne of 9 June 1787 (Dawes, 1787j: 268r). Meanwhile, Hunter (1793: 26–28) included a table of their geographic locations, as well as the prevailing weather conditions, on the leg from Tenerife to Rio de Janeiro. Since most journals of First Fleet officers are now readily accessible online (Project Gutenberg Australia, 2019), we consulted the journals of John Hunter (1793), William Bradley (1786–1792), David Collins (1798), Arthur Phillip (1790a,b), Arthur Bowes Smyth (Choat, 2020), John White (1790) and Daniel Southwell (Bladen, 1893). We also perused William Dawes' correspondence contained in the Board of Longitude papers. We thus collected a total of 287 unique geographic location references, which we have made publicly available.[24]

Bradley included a number of charts of the *Sirius*' detailed itinerary in his journal based on longitude and latitude measurements. Here, we have reproduced his charts of the first leg of the voyage from England across the Equator in the eastern Atlantic Ocean (see Figure 18) and an overview chart of the southern hemisphere section, from a location just north of the Equator to Rio de Janeiro, Table Bay and New South Wales (see Figure 19). Combined with Figure 10, this selection of charts offers detailed insights into the movements of the *Sirius* during her long voyage to the new settlement (for a modern representation, see Gergis et al., 2010, their Figure 3). Note, in particular, the short-term deviations from the expected direct routes, which were due to adverse winds (indicated by wind arrows) or their absence, in the doldrums. In addition, some deviations were owing to currents, mostly in the Atlantic Ocean where the convoy was subject to strong West→East currents south of the Equator (e.g., Hunter, 1793).

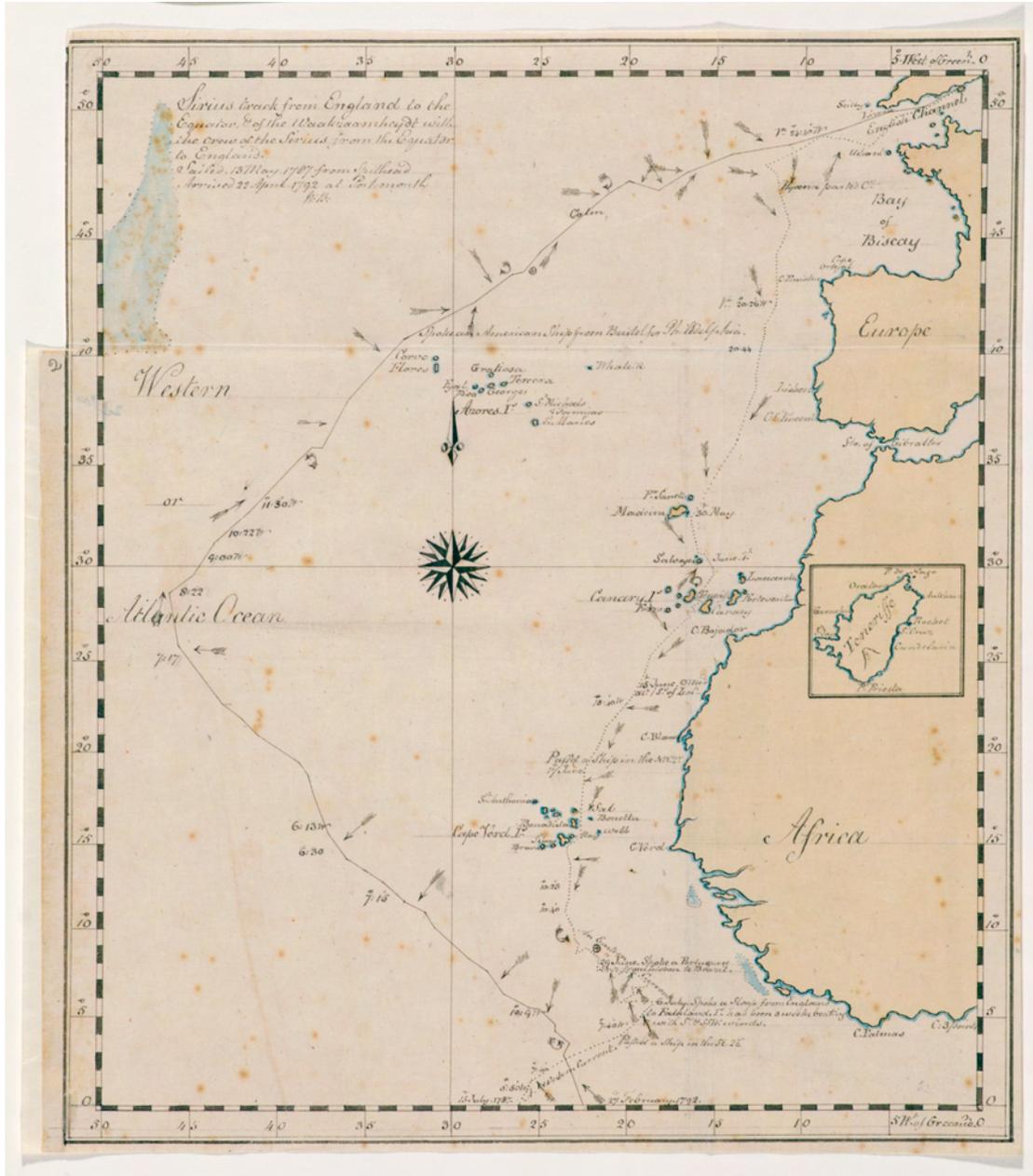

**Figure 18**: "*Sirius* track from England to the Equator, & of the *Waakzaamheydt* with the crew of the *Sirius* from the Equator to England". The *Sirius*' itinerary is indicated by the dotted line. Chart from William Bradley's journal, *A Voyage to New South Wales*, ca. 1802 (Mitchell Library, State Library of New South Wales; Ref. 404927).

It is interesting to note the discrepancies among the different observers, particularly as regards their crossing of the Equator and the Tropic of Cancer (in the northern hemisphere). While Collins, Tench, Hunter and White noted that they crossed the Equator on 14 July 1787, Bradley and King refer to the 15th and Phillip cites 5 July 1787 (this may be a transcription error). One may think that this discrepancy could be related to the use of the astronomical (or nautical) versus the civil day, with the astronomical day starting at 12 noon rather than at 12 midnight. However, that explanation seems unlikely. Among those who cited their crossing of the Equator, Hunter and Bradley were the most experienced observers, and it is likely that the others followed their lead.[25] Yet, Hunter indicates that the Equator was crossed during the "evening" of 14 July, while Bradley and King recorded specific times of, respectively, 5 pm and 8 pm, but on the 15th. These timings are well matched, and so perhaps the discrepancy is due to the officers having completed their journal entries from memory at a later date. Similarly, both Collins and Bradley note that the Tropic of Cancer was crossed on 15 June 1787, but White's record refers to 10 June. We suspect that this latter record is a mistake given the accuracy of Bradley's astronomical observations and of Collins' careful chronicling of the convoy's itinerary.

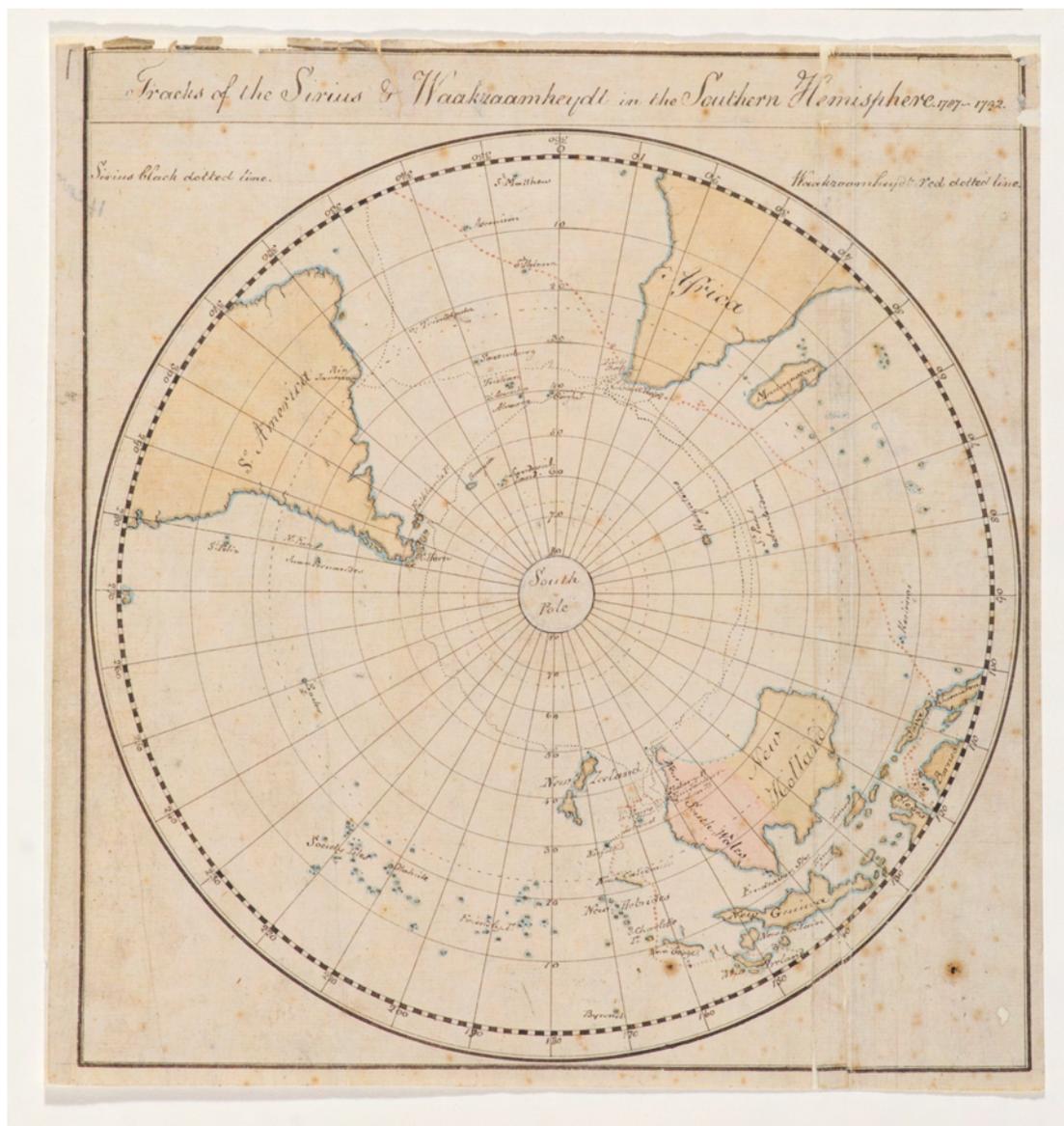

**Figure 19**: "Tracks of the *Sirius* & *Waakzaamheydt* in the Southern Hemisphere, 1787–1792". The *Sirius*' itinerary is indicated by the thin dotted line. Note the distinction between 'New Holland' and 'New South Wales' (including its size compared with today's state by the same name) on the map of Australia. Chart from William Bradley's journal, *A Voyage to New South Wales*, ca. 1802 (Mitchell Library, State Library of New South Wales; Ref. 404927). [For a high-resolution version, see http://archival.sl.nsw.gov.au/Details/archive/110314967.]

Our complete data set of 287 entries includes a total of 37 longitude and latitude determinations of well-defined geographic locations from First Fleet journals. To these, we added two of Cook's position determinations of features on the Tasmanian coast (Bradley, 1787–1792), given that they were referenced by the First Fleet's chroniclers. Here, we will make an attempt at comparing the accuracy and precision of the contemporary measurements with modern values. Our complete data set, as well as a direct comparison with modern (2020) *Google Maps*® locations, is provided below; 'Time-keeper' refers to K1 measurements.

2. *Deserter Islands (Ilhas Desertas, Desertas Islands)*:
   a. Northernmost Deserter Island

| *Latitude* | *Longitude* | *Means* | *Observer* | *Modern Latitude* | *Modern Longitude* |
|---|---|---|---|---|---|
| 32º27'N | 16º35'W | Time-keeper | Bradley | 32º35'N | 16º33'W |

   b. Southeasternmost Deserter Island

| *Latitude* | *Longitude* | *Means* | *Observer* | *Modern Latitude* | *Modern Longitude* |
|---|---|---|---|---|---|
| 32º29'N | 16º38'W | Time-keeper | Hunter | 32º25'N | 16º28'W |
| 32º28'N | | | Tench | | |

2. *Salvage Islands*:
   c. Mean

| *Latitude* | *Longitude* | *Means* | *Observer* | *Modern Latitude* | *Modern Longitude* |
|---|---|---|---|---|---|
| 30º12'N | 15º56'W | Time-keeper | Bradley | 30º08'N | 15º52'W |
| 30°13'N | 15º56'W | Time-keeper | King | | |
| 30º12'N | 15º53'W | Time-keeper | Hunter | | |

   d. Great Salvage, eastern side

| *Latitude* | *Longitude* | *Means* | *Observer* | *Modern Latitude* | *Modern Longitude* |
|---|---|---|---|---|---|
| 30º12'N | 15º39'W | | Tench | 30º08'N | 15º52'W |

3. *Tenerife (Canary Islands)*:
   a. Santa Cruz bay

| *Latitude* | *Longitude* | *Means* | *Observer* | *Modern Latitude* | *Modern Longitude* |
|---|---|---|---|---|---|
| 28º30'N | 16º16'W | Time-keeper | Bradley | 28º29'N | 16º14'W |
| 28º29'N | 16º18'W | Time-keeper | Hunter | | |
| | 16º18'W | Time-keeper | King | | |
| | 16º18'W | | Dawes | | |

   b. Santa Cruz (old town)

| *Latitude* | *Longitude* | *Means* | *Observer* | *Modern Latitude* | *Modern Longitude* |
|---|---|---|---|---|---|
| 28º28'N | 16º18'W | | Tench | 28º28'N | 16º15'W |

   c. Teide volcano (peak)

| *Latitude* | *Longitude* | *Means* | *Observer* | *Modern Latitude* | *Modern Longitude* |
|---|---|---|---|---|---|
| 28º18'N | 16º31'W | | Bradley | 28º16'N | 16º39'W |

4. *Cape Verde Islands*:
   a. Sal, northern end

| *Latitude* | *Longitude* | *Means* | *Observer* | *Modern Latitude* | *Modern Longitude* |
|---|---|---|---|---|---|
| 16º48'N | 23º03'W | Time-keeper | Hunter | 16º51'N | 22º55'W |
| 16º50'N | 23º02'W | Time-keeper | Bradley | | |

   b. Sal, southern end

| *Latitude* | *Longitude* | *Means* | *Observer* | *Modern Latitude* | *Modern Longitude* |
|---|---|---|---|---|---|
| 16º40'N | 23º02'W | Time-keeper | Bradley | 16º35'N | 22º55'W |
| 16º40'N | 23º05'W | | Tench | | |
| 16º15'N | 22º51'W | Time-keeper | King | | |

c. Bona Vista (Boa Vista), northern end

| Latitude | Longitude | Means | Observer | Modern Latitude | Modern Longitude |
|---|---|---|---|---|---|
| 16º13'N | 22º51'W | Time-keeper | Hunter | 16º14'N | 22º47'W |

d. Bona Vista (Boa Vista), northern end

| Latitude | Longitude | Means | Observer | Modern Latitude | Modern Longitude |
|---|---|---|---|---|---|
| 15º59'N | 23º02'W | Time-keeper | Bradley | 15º58'N | 22º48'W |
| 15º52'N | 23º08'W |  | Tench |  |  |
| 16º00'N | 22º51'W | Time-keeper | Hunter |  |  |

e. Isle of May (Maio), southern end

| Latitude | Longitude | Means | Observer | Modern Latitude | Modern Longitude |
|---|---|---|---|---|---|
| 15º11'N | 23º26'W |  | Tench | 15º07'N | 23º10'W |

f. St. Jago, Port Praya fort

| Latitude | Longitude | Means | Observer | Modern Latitude | Modern Longitude |
|---|---|---|---|---|---|
|  | 23º37'W |  | Tench | 14º44'N | 23º31'W |

5. *Brazil*:
   a. Cape Frio (Cabo Frio)

| Latitude | Longitude | Means | Observer | Modern Latitude | Modern Longitude |
|---|---|---|---|---|---|
| 23º05'S | 41º40'W |  | Tench, Dawes | 22°47'S | 41°56'W |
| 23º00'S | 41º44'W | Time-keeper | Bradley |  |  |
| 22º58'S | 41º40'W | Time-keeper | Hunter (*) |  |  |

(*) "It will appear hereafter that we had not the true rate of the watch, and consequently that [this] longitude is not correct" (Hunter, 1793: 17).

b. Rio de Janeiro

| Latitude | Longitude | Means | Observer | Modern Latitude | Modern Longitude |
|---|---|---|---|---|---|
| 22°54'S | 42°40'W |  | Bowes Smyth | 22°54'S | 43°11'W |
| 22°54'S | 42°44'W | Time-keeper | Hunter (*) |  |  |
| 22°54'S | 43°19'W |  | White |  |  |
| 22°54'S | 43°11'W | Lunar distances | Hunter |  |  |
| 22°54'S | 43°33'W | Lunar distances | Bradley |  |  |
| 22°54'S | 43°19'W | Jupiter's 3rd moon | Dawes, Portuguese astronomers |  |  |

(*) See the note to Table 5a.

c. Enchados (Enxadas) Island, Rio de Janeiro

| Latitude | Longitude | Means | Observer | Modern Latitude | Modern Longitude |
|---|---|---|---|---|---|
|  | 43°21'W | Time-keeper | King | 22°53'S | 43°11'W |

6. *Final leg: from Table Bay to Botany Bay*:
   a. Table Bay

| Latitude | Longitude | Means | Observer | Modern Latitude | Modern Longitude |
|---|---|---|---|---|---|
| 34°22'S | 18°45'E |  | Bowes Smyth | 33°53'S | 18°27'E |

b. Cape Town (old town)

| Latitude | Longitude | Means | Observer | Modern Latitude | Modern Longitude |
|---|---|---|---|---|---|
| 33°55'S | 18°24'30"E | Lunar distances | Hunter | 33°54'S | 18°25'E |
| | 18°23' 55"E | Lunar distances | Collins | | |

c. Mewstone

| Latitude | Longitude | Means | Observer | Modern Latitude | Modern Longitude |
|---|---|---|---|---|---|
| 43°48'S | 146°25'E | | Tench, Cook | 43°44'S | 146°22'E |

d. Eddystone and Swilly Rock[26]

| Latitude | Longitude | Feature | Observer | Modern Latitude | Modern Longitude |
|---|---|---|---|---|---|
| 43°54'S | 147°09'E | Eddystone | Tench | 43°47'S | 147°01'E |
| 43°53'S | 147°09'E | Eddystone | Cook | | |
| 43°54'S | 147°03'E | Swilly Rock | Tench | 43°47'S | 147°01'E |

It is interesting to compare the accuracy of the contemporary longitude and latitude determinations with current-best estimates. On the whole, it transpires that the First Fleet's measurements were systematically offset by 2.4' towards larger longitude differences with respect to the Greenwich meridian and by 3.6' in latitude (offset towards northern latitudes), respectively, but with large standard deviations of 11.5' and 7.0', respectively. The systematic offset in longitude is reminiscent of a similar (but larger) systematic offset *in the same sense* recorded by Dawes, Hunter and Bradley for the geographic location of Dawes' observatory in the new colony compared with modern measurements (de Grijs and Jacob, 2021; and references therein). A similar discrepancy was found for the location of the semi-permanent observatory on the shores of Botany Bay established by the Lapérouse expedition. In de Grijs and Jacob (2021) we concluded that these systematic offsets are most likely owing to problems related to the accuracy of the contemporary almanac tables.

We can directly convert the systematic angular differences just cited to time differences. In this context, we are particularly interested in the longitude difference: the systematic offset of 2.4' corresponds to a time discrepancy of 9.6 seconds of time. The longitude offset for Dawes' observatory amounted to almost 29 seconds of time (de Grijs and Jacob, 2021). Therefore, at the latitude of Sydney, 33° 52' 08" South, the corresponding average offset in longitude achieved during the First Fleet's voyage would be less than 4 km. Consideration of this accuracy, achieved during the First Fleet's voyage *at sea and on shore*, then begs the question as to why the longitude determination of Dawes' observatory *on shore only* was off by as much as 11 km.

Whereas some contemporary position determinations rivalled the accuracy and precision of modern measurements, particularly those obtained in well-known locations such as Santa Cruz de Tenerife, Rio de Janeiro and Cape Town, shipboard measurements of distant landmarks often incurred significant uncertainties. The most significant differences are found for some of the Cape Verde islands, for Cape Frio on the Brazilian coast and, surprisingly, for the geographic location of Table Bay. In fact, the Table Bay measurement recorded by Bowes Smyth is one of the most discrepant determinations in our database: "13 October 1787 … Anchored at Cape of Good Hope. Lat[itude] 34°22'S, long[itude] 18°45'E" (Choat, 2020: 53). This location corresponds to the southeastern entrance to False Bay, south of Cape Town and just off the coast of present-day Pringle Bay village in the Western Cape. It is clear from Bradley's maps (available online from the State Library of New South Wales; see the caption of Figure 19), however, that the convoy anchored in Table Bay (Bradley, 1786–1792). Bowes Smyth's contemporary record may have been incorrectly copied, given that he was assigned to the *Lady Penrhyn* and thus without a direct means of geographic position determination himself. We suspect that some of the offshore measurements of island features may reflect the ships' locations rather than the island features themselves, but this cannot be verified based on the contemporary record.

Among the more experienced observers for whom we have sufficient numbers of measurements (Hunter: 9; Bradley: 9; and Tench: 10), Hunter's longitude determinations were most accurate (with a mean difference compared with modern measurements closest to zero). Hunter's nine longitude measurements deviate, on average, by –1.4', compared with longitude differences of +4.2' and +4.1' resulting from, respectively, Bradley's and Tench's observations. The corresponding standard deviations for all three observers span a narrow range from 9.7' to 11.4'. Latitude determinations have always been easier and were (usually) more accurate and precise (corresponding to smaller standard deviations) than longitude determinations. On average, with respect to modern measurements, our three most prolific observers determined their latitudes to within 1.9' (Hunter), 2.0' (Bradley) and 2.7' (Tench), with all observers reporting latitudes offset from modern values to the north. Hunter's precision exceeded that of his colleagues, with a standard deviation of 3.9' for his set of latitude determinations, compared with 5.1' and 6.8' for Bradley and Tench, respectively.

Whereas Dawes was known to be a careful astronomer, his measurements were good but not better than those of Hunter. It is perhaps surprising that Dawes only recorded a small number of longitude determinations. Nevertheless, it is clear from our careful perusal of his letters to Maskelyne that he obtained numerous lunar distance measurements. However, it appears that he did not regularly convert those observations into the convoy's longitude at the time of observation. This would have required lengthy calculations.

With Dawes occupied on the First Fleet's vanguard ship, *Supply*, the remainder of the convoy under Hunter's command was clearly in safe hands. From our modern perspective, it is challenging to adequately visualise Captain Phillip's remarkable leadership. Given the navigational means available at the time, it was indeed a major achievement to direct an entire convoy of 11 ships on a 24,000 km, 8-month voyage half-way around the world without the loss of a single ship, and to ensure that all ships arrived safely within a few days of one another with minimum loss of life despite the cramped living quarters. They could not have done it without a firm grasp of the principles of practical astronomy.

**6 NOTES**

[1] All ships, not only the food and supply store-ships, were stocked to the brim with provisions, agricultural and camp equipment, clothing for the convicts, baggage and numerous other items (Collins, 1798).

[2] A normal human pregnancy, from ovulation to natural childbirth, lasts 268 ± 9 days (standard deviation; Jukic et al., 2013: Table 1), which implies that all children born in transit were most likely conceived prior to the First Fleet's departure. Statistics of births in the colony are sparse for the first few years. In a letter of 12 February 1790 to Lord Sydney, Phillip (1790b: 144) states, "As near two years have now passed since we first landed in this country ... Fifty-nine children have been born in the above time." Given this timeframe this number does not tell us whether any children were conceived during the voyage. Anecdotal evidence suggests that a number of births occurred in the first few months after the convoy's arrival, however (e.g., National Library of Australia, n.d.; *Australian Town and Country Journal*, 1883; *Nepean Times*, 1883; *Daily Telegraph*, 1888).

[3] In de Grijs and Jacob (2021) we already included the portraits of William Dawes, Arthur Phillip, John Hunter, David Collins and Watkin Tench. Here, we show the portraits of the remaining individuals of note discussed in this article, if available in the public domain.

[4] The small town of Wallabadah in the New England area of present-day New South Wales is home to the First Fleet memorial garden. The local visitors' centre maintains a list of those known to have sailed on the First Fleet, which currently contains 1186 names. However, the list does not fully tally with the names engraved on stone tablets scattered around the garden (Liverpool Plains Shire Council, 2020; N. Robertson, personal communication).

[5] See also https://history.cass.anu.edu.au/centres/ncb/first-fleet-ships-and-passengers

[6] References to 'Nicolas' as William Dawes' middle name are found only at *Wikipedia*/*WikiTree*—https://en.wikipedia.org/wiki/William_Dawes_(British_Marines_officer), https://www.wikitree.com/wiki/Dawes-1222—but not anywhere else.

[7] This was the Atlantic season's fourth hurricane, known as the 1871 Santa Juana hurricane (https://en.wikipedia.org/wiki/1871_Atlantic_hurricane_season).

[8] https://cudl.lib.cam.ac.uk/view/MS-RGO-00014-00048/509

[9] English parish registers usually only recorded baptisms rather than births. In England, civil registration started in 1837.

[10] Graduates from the Royal Naval Academy were often appointed as midshipman-by-order or 'midshipman ordinary' to make a clear distinction from the midshipmen who had been trained on board. The latter were paid higher wages (Lewis, 1939: 217).

[11] Wright (1927) reported that, on 27 September 1926, the Adjutant-General of the Royal Marines at the Admiralty in London responded to a request from the Royal Australian Historical Society about Dawes' career, confirming that "[the] only particulars shown in official records are as follows: Granted Commission as 2nd Lieutenant, Royal Marines, 2nd September, 1779; …" This is an odd response from the official body, given that the Marines did not become *Royal* until 1802.

[12] In the British Royal Navy's rating system, from the 1720s a third-rate warship carried 64 to 80 guns, typically on two gun decks.

[13] Maskelyne had convinced Dawes that taking on the role of the voyage's astronomer might lead to further career opportunities, and therefore Dawes did not seek financial compensation for the appointment (Clarke, 2015).

[14] Laurie (1988) suggests that Dawes may have approached Maskelyne directly, given his father's standing in the Portsmouth Ordnance Office and the elder Dawes' acquaintance with the Royal Naval Academy staff and with the family of Lord Hood, previously Commissioner at the dockyard and then-Commander-in-Chief at Portsmouth. This seems less likely than Bayly's approach, however.

[15] Dawes' relatively quick acceptance of his assignment on the *Sirius* was likely driven, at least in part, by the fact that he was near the top of the Marines' rotation roster, which implies that he would have shortly been allocated to another vessel if he had remained in Portsmouth (Bayly, 1786).

[16] Dawes was required to seek Maskelyne's approval for any adjustments to the instruments. At times, this caused strained relations between the men, since Dawes (based at Portsmouth) was left feeling as if he was being micromanaged from afar (Saunders, 1990: 46, 52).

[17] http://www.visitsydneyaustralia.com.au/william-dawes.html

[18] Dawes' name and achievements are conspicuously absent from Phillip's (1790a) account of the voyage to New South Wales.

[19] This watch had been taken on Cook's second and third voyages of discovery to the Pacific in the care of his astronomer, William Wales (Howse and Hutchinson, 1969: 139–140).

[20] In good Admiralty tradition, scientific instruments were usually stored near a ship's supply of bread, since bread was kept in the driest place on the ship (King, 1779; Saunders, 1990: 67).

[21] However, Hunter (1793: 17) cast doubt on the party's longitude determinations: "It will appear hereafter that we had not the true rate of the watch, and consequently that the above longitude is not correct".

[22] Note that Phillip only mentions King as having transferred with him to the *Supply* (Phillip, 1790a: 50); he does not mention Dawes.

[23] Hunter (1793: 38–39) noted that "The latitudes and longitudes of the different points or capes, seem to have been very correctly determined by Captains Cook and [Tobias] Furneaux, when they were here; … Such observations as we had an opportunity of making near this coast, agree very well ..."

[24] See http://astro-expat.info/Data/First_Fleet_positions.html. This webpage has been saved for posterity at http://web.archive.org/web/20201116085541/http://astro-expat.info/Data/First_Fleet_positions.html by the *Wayback Machine*.

[25] As noted in our online database (note 24), White—who was based on the *Charlotte* without access to a chronometer or astronomical equipment—indicates time and time again that his geographic locations were obtained from measurements displayed by the *Sirius*.

[26] The modern coordinates provided for 'Swilly Rock' are only approximate. Matthew Flinders (1814: Section IV, part I) identified Swilly Rock as the westernmost of a pair of islets: "They had, a little before, passed two cliffy islets lying to seaward; of which the westernmost (*Swilly* of Furneaux) is like *Pedra Blanca* near the coast of China; the easternmost (*Eddystone* of Cook) resembles an awkward tower, and is about sixteen miles from the mainland."

## 7 ACKNOWLEDGEMENTS

We are grateful to the library staff at the Special Collections desk of the Mitchell Library, State Library of New South Wales (Sydney, Australia), and the National Library of Australia (Canberra). We also thank Nikki Robertson of the Liverpool Plains Shire Council visitor information centre for providing background information about (and a personal guided tour of) the First Fleet memorial garden in Wallabadah. We also acknowledge a number of valuable suggestions by the reviewers that made this paper more interesting.